\documentclass[%
reprint,
superscriptaddress,
 amsmath,amssymb,
 aps,
]{revtex4-2}

\usepackage{amsmath}
\let\oldAA\AA
\renewcommand{\AA}{\text{\normalfont\oldAA}}
\usepackage{xcolor}
\usepackage{multirow}

\usepackage{booktabs}
\usepackage{lipsum}
\usepackage{graphicx}
\usepackage{dcolumn}
\usepackage{bm}
\usepackage{amsmath}
\usepackage{amssymb}
\usepackage{dutchcal}
\usepackage{float}
\begin{document}

\title{Excitonic Insulator to Superconductor Phase Transition in Ultra-Compressed Helium}

\author{Cong Liu}
\affiliation{Departament de Física, Universitat Politècnica de Catalunya, Campus Nord B4-B5, Barcelona 08034, Spain}

\author{Ion Errea}
\affiliation{Fisika Aplikatua Saila, Gipuzkoako Ingeniaritza Eskola, University of the Basque Country (UPV/EHU), Europa Plaza 1, 20018 Donostia/San Sebastián, Spain; Centro de Física de Materiales (CSIC-UPV/EHU), Manuel de Lardizabal pasealekua 5, 20018 Donostia/San Sebastián, Spain; Donostia International Physics Center (DIPC), Manuel de Lardizabal pasealekua 4, 20018 Donostia/San Sebastián, Spain}

\author{Chris Pickard}
\affiliation{Department of Materials Science and
Metallurgy, University of Cambridge, Cambridge CB30FS,
United Kingdom; Advanced Institute for Materials Research, Tohoku
University, Sendai 980-8577, Japan}

\author{Lewis J. Conway}
\affiliation{Department of Materials Science and
Metallurgy, University of Cambridge, Cambridge CB30FS,
United Kingdom; Advanced Institute for Materials Research, Tohoku
University, Sendai 980-8577, Japan}

\author{Bartomeu Monserrat}
\affiliation{Cavendish Laboratory, University of
Cambridge, Cambridge CB30HE, United Kingdom.; Department of Materials Science and Metallurgy, University of Cambridge, Cambridge CB30FS, United Kingdom}

\author{Yue-Wen Fang}
\affiliation{Fisika Aplikatua Saila, Gipuzkoako Ingeniaritza Eskola, University of the Basque Country (UPV/EHU), Europa Plaza 1, 20018 Donostia/San Sebastián, Spain; Centro de Física de Materiales (CSIC-UPV/EHU), Manuel de Lardizabal pasealekua 5, 20018 Donostia/San Sebastián, Spain}

\author{Chi Ding}
\affiliation{National Laboratory of Solid State Microstructures, School of Physics and
Collaborative Innovation Center of Advanced Microstructures, Nanjing University, Nanjing 210093, China}

\author{Qing Lu}
\affiliation{National Laboratory of Solid State Microstructures, School of Physics and
Collaborative Innovation Center of Advanced Microstructures, Nanjing University, Nanjing 210093, China}

\author{Jian Sun}
\affiliation{National Laboratory of Solid State Microstructures, School of Physics and
Collaborative Innovation Center of Advanced Microstructures, Nanjing University, Nanjing 210093, China}

\author{Jordi Boronat}
\affiliation{Departament de Física, Universitat Politècnica de Catalunya, Campus Nord B4-B5, Barcelona 08034, Spain}

\author{Claudio Cazorla}
\email{Corresponding author: claudio.cazorla@upc.edu}
\affiliation{Departament de Física, Universitat Politècnica de Catalunya, Campus Nord B4-B5, Barcelona 08034, Spain}

\begin{abstract}
Helium, the second most abundant element in the universe, exhibits an extremely large electronic band gap of about $20$~eV at low pressures ($\le 0.1$~GPa). While the metallization pressure of hcp helium has been accurately predicted, thus far little attention has been paid to the specific mechanisms driving the band-gap closure and electronic properties of this quantum crystal in the terapascal regime (1~TPa $= 1,000$~GPa). Here, we employ state-of-the-art density functional theory and many-body perturbation theory calculations to fill up this knowledge gap. It is found that prior to reaching metallicity bulk solid helium becomes an excitonic insulator (EI), an exotic state of matter typically observed in low-dimensional systems in which electrostatically bound electron-hole pairs form spontaneously. Furthermore, it is shown that electron-phonon coupling (EPC) is significantly enhanced across the EI to metal phase transition as signaled by prominent phonon softening and giant EPC strength values ($\lambda \sim 10$--$100$) estimated at specific reciprocal space points. Accordingly, we predict metallic helium to be a superconductor with a critical temperature of $\approx 30$~K at $20$~TPa and of $\approx 100$~K at $100$~TPa. These unforeseen phenomena have important consequences on the elastic, thermodynamic and transport properties of metallic helium hence may be critical for improving our fundamental understanding and modelling of celestial bodies.
\end{abstract}
\maketitle

In their final evolution stage, most stars in the Universe become white dwarfs (WDs)  consisting of a mixture of helium, carbon, and oxygen atoms immersed in a sea of electrons. In small-radius WDs, fusion reactions beyond helium hardly occur hence their chemical composition is practically monoelemental. In the interior of WDs, pressure may reach values billions of times higher than that in the Earth's surface ($\sim$~10,000~GPa~=~10~TPa), which currently are not accessible in experiments. Thus, theoretical modelling of light materials under extreme compression conditions, and in particular of helium, turns out to be critical for probing the interior of WDs and comprehend their physico-chemical evolution.

\begin{figure*}[t]
\centering
\includegraphics[width=0.8\textwidth]{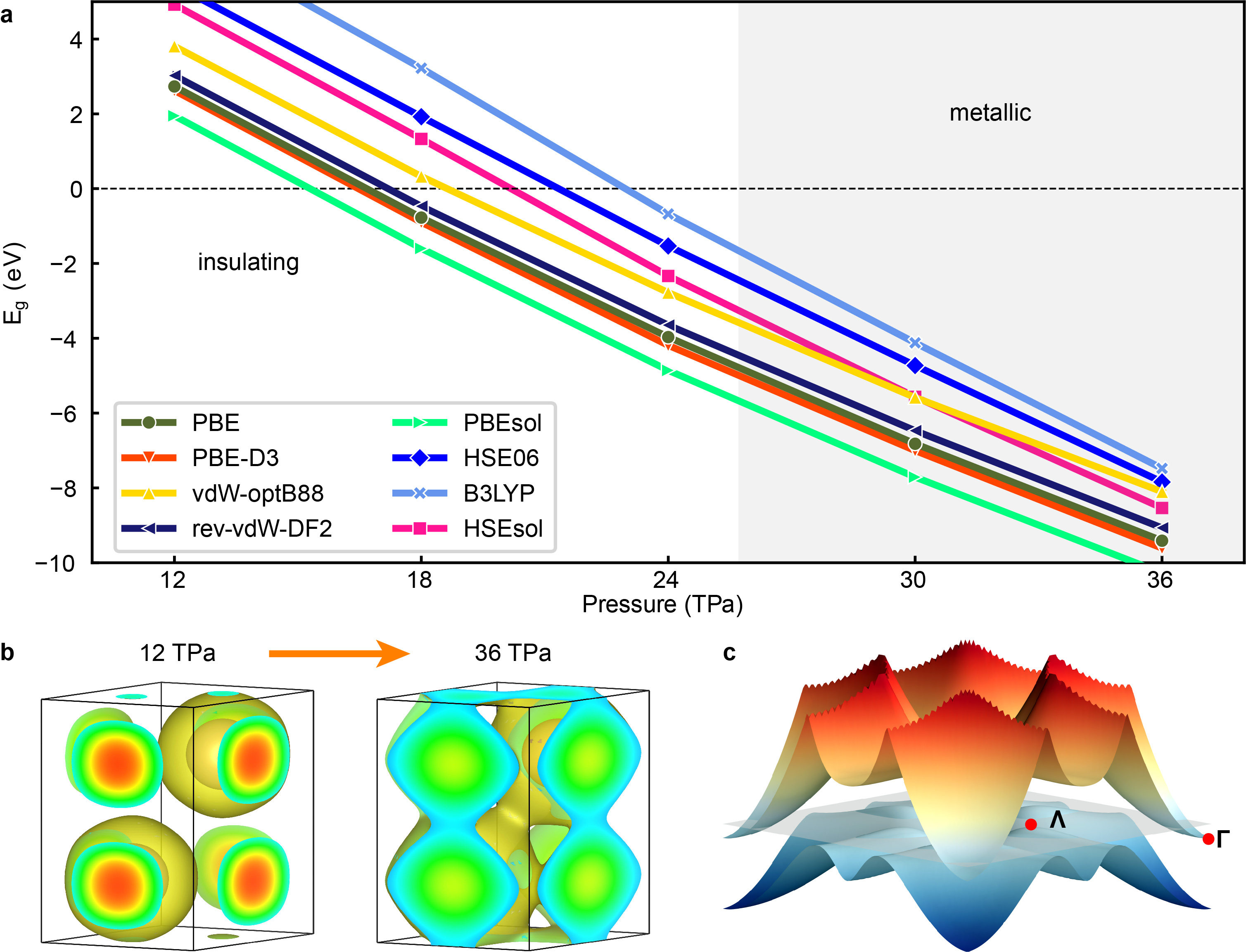}
\caption{\textbf{DFT benchmarking for the metallization pressure of hcp $^{4}$He.} \textbf{a}~Electronic band gap, $E_{g}$, expressed as a function of pressure and calculated with different DFT functionals. Negative $E_{g}$ values indicate overlapping between the VBM and CBM levels. The grey region indicates the stability range of metallic helium as calculated with QMC methods \cite{khairallahFirstPrinciplesStudiesMetallization2008a}. \textbf{b}~Electronic localization function (ELF isosurface~=~$0.8$, yellow) of solid helium at $12$ and $36$~TPa in a red-green-blue color scale with red denoting high electronic density and blue low electronic density. \textbf{c}~Lowest conduction (red) and highest valence (blue) bands expressed as a function of reciprocal wave vector in the $k_z$ = 0 plane; the grey and transparent plane represents the Fermi surface.}
\label{fig:benchmark}
\end{figure*}

Highly accurate diffusion Monte Carlo (DMC) calculations predict solid helium to become a metal in the hexagonal closed packed (hcp) phase at a pressure of $25.7$~TPa \cite{khairallahFirstPrinciplesStudiesMetallization2008a}. By considering zero-point energy and electron-phonon coupling effects estimated with density functional theory (DFT) methods, such a metallization pressure increases up to $32.9$~TPa at $T = 0$~K \cite{monserratElectronPhononCouplingMetallization2014a}. Both experimental and theoretical studies have shown that the valence-band maximum (VBM) of $^{4}$He appears on the line joining the reciprocal lattice points $\Gamma$~$(0,0,0)$ and $M$~$(q,0,0)$, while the conduction-band minimum (CBM) is located at the $\Gamma$ point \cite{monserratElectronPhononCouplingMetallization2014a,maoElectronicStructureCrystalline2010}. Thus, the band gap of solid helium is indirect and according to previous DFT calculations the overlap between the conduction and valence bands when approaching metallization is characteristic of a semimetal (i.e., the density of electronic states at the Fermi level is negligible) \cite{monserratElectronPhononCouplingMetallization2014a}. Meanwhile, the lattice phonons involving atomic displacements perpendicular to the hcp basal plane strongly couple to the electronic bands and at very high pressure these drive the widening of the band gap \cite{monserratElectronPhononCouplingMetallization2014a}.

A detailed understanding of the electronic band structure properties of this archetypal quantum crystal \cite{cazorla17}, however, is still lacking. First, about half a century ago the existence of an exotic insulating phase called ``excitonic insulator'' (EI) was predicted in which electrons and holes spontaneously form bound pairs called excitons \cite{jeromeExcitonicInsulator1967}. The EI phase could be stabilized at sufficiently low temperatures in semiconductors with tinny band gaps or semimetals with very small band overlaps. Recently, experimental EI fingerprints have been reported for low-dimensional transition metal dichalcogenide structures exhibiting small band gaps \cite{ma21,jia22}; however, stabilization of a bulk EI state remains elusive. Owing to its semiconductor nature, absence of structural transformations and marked quantum character, ultra-compressed $^{4}$He appears to be an excellent candidate in which a bulk EI state could emerge and genuine quantum many-body phenomena like high-temperature excitonic superconductivity and BEC--BCS crossover might occur \cite{novoselov14,bronold06}. Is possibly solid helium a bulk EI in the TPa regime? And second, the substantial electron-phonon coupling and semimetal Fermi surface previously disclosed in solid helium suggest the possibility of superconductivity in this quantum crystal upon band-gap closure. Is metallic helium a superconductor? If so, what are the underlying physical mechanisms and corresponding critical temperature? Besides their fundamental interest, answering to these questions may have major consequences in the fields of planetary science and astrophysics since this new knowledge could improve our understanding of the thermal and chemical evolution of small-radius WDs \cite{wd1,wd2}.

In this Letter, we employ theoretical first-principles approaches based on DFT and many-body perturbation theory to advance knowledge on the electronic, elastic and superconductor properties of solid helium in the TPa regime. Our main finding is an unprecedented bulk excitonic insulator to superconductor phase transition driven by pressure in which the superconductor state can reach a critical temperature of $\approx 100$~K under a compression of $100$~TPa. It is worth noting that an exhaustive random sampling of the structural space of solid helium was performed at $P = 100$~TPa (AIRSS \cite{airss1,airss2}), with the finding that the hcp phase imperturbably remained the ground state (Methods).

\begin{figure*}[htp]
\centering
\includegraphics[width=0.9\textwidth]{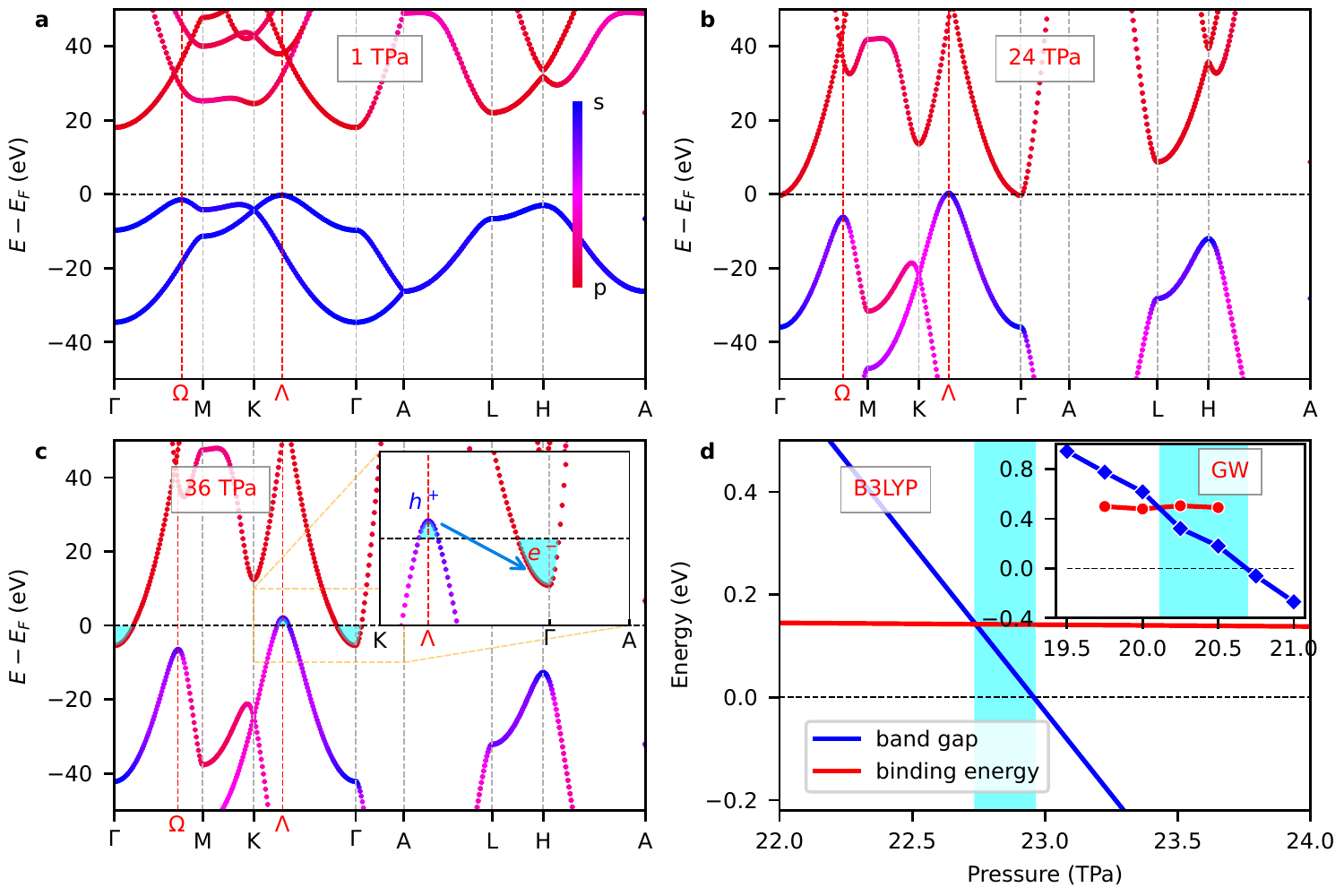}
\caption{\textbf{Band-gap closure and emergence of the excitonic insulator state in hcp $^{4}$He.} \textbf{a-c}~Evolution of the electronic band structure under compression calculated with the hybrid B3LYP functional. A red-magenta-blue color scale is employed for representing the orbital character of the relevant electronic bands with blue for $s$- and red for $p$-like. The $\Lambda$ and $\Omega$ reciprocal space points indicate the location of the primary and secondary VBM. The inset highlights the migration of electrons ($e^{-}$) and formation of holes ($h^+$) along the reciprocal space line $K$-$\Lambda$-$\Gamma$ near the Fermi surface. \textbf{d}~Comparison of the exciton binding energy, $E_{bind}$, and band gap calculated with B3LYP DFT and many-body perturbation theory within the GW approximation. The blue regions indicate the pressure range in which the sufficient condition for spontaneous formation of excitons is fulfilled.}
\label{fig:orbit}
\end{figure*}

We started by benchmarking different families of DFT functionals (i.e., semi-local, van der Waals corrected and hybrid) \cite{cazorla17} against the metallization pressure of solid helium calculated with DMC methods, which amounts to $25.7$~TPa (Fig.\ref{fig:benchmark}a). In all the analyzed cases, the band gap decreases almost linearly under increasing pressure due to the steady enhancement of electronic delocalization among neighbouring atoms (Fig.\ref{fig:benchmark}b). The semi-local PBE functional predicts a metallization pressure of $17$~TPa, in consistent agreement with previous computational studies \cite{khairallahFirstPrinciplesStudiesMetallization2008a,monserratElectronPhononCouplingMetallization2014a}. Meanwhile, the hybrid functional B3LYP performs the best in comparison to the DMC benchmark by providing a metallization pressure of $\approx 24$~TPa (Fig.\ref{fig:benchmark}a and Fig.\ref{fig:orbit_compression}). Van der Waals corrections turn out to be practically negligible in the TPa regime (e.g., the two PBE and PBE-D3 curves practically overlap each with the other) due to the dominant role of interatomic repulsive interactions at short distances \cite{cazorla15}. Based on these results, we adopted the hybrid functional B3LYP for our subsequent analysis of the electronic band structure of solid helium. 

Unlike atomic hydrogen, hcp $^{4}$He presents an indirect band gap with the VBM located at the reciprocal point $\Lambda$ and the CBM at the center of the Brillouin zone ($\Gamma$ point, Fig.\ref{fig:benchmark}c), in consistent agreement with previous DFT calculations and experiments \cite{maoElectronicStructureCrystalline2010, monserratElectronPhononCouplingMetallization2014a}. It is noted that the direct band gap at the $\Lambda$ point actually increases under compression (Fig.~\ref{fig:direct_bandgap}). Interestingly, when the energy gap between the VBM and CBM levels disappears a semimetal state characterized by an almost negligible density of states at the Fermi level emerges due to the fact that no additional electronic bands cross the Fermi surface (Fig.\ref{fig:orbit}a--c). At $1$~TPa, the VBM consists exclusively of $s$-like orbitals while the CBM exhibits full $p$-like character (Fig.\ref{fig:orbit}a). Upon further compression, the VBM presents increasingly larger hybridization between $s$ and $p$-like orbitals while the CBM conserves its pure $p$-like character (Fig.\ref{fig:orbit}b). At pressures higher than $24$~TPa, electrons from the VBM at the $\Lambda$ point are transferred to the CBM minimum at $\Gamma$ in order to lower their energy, thus rendering a $p$-type semimetal system (Fig.\ref{fig:orbit}c). 

The continuous pressure-driven closure of the band gap and subsequent stabilization of a semimetal state in hcp $^{4}$He, suggest the possibility of spontaneous formation of excitons with finite momentum $|\bf{\rm{q}}|=\Lambda$--$\Gamma$ at low temperatures. An exciton is a bound state formed by an excited electron ($e^{-}$) in the conduction band and a hole ($h^{+}$) in the valence band that interact through attractive Coulomb forces. In narrow-gap semiconductors, a sufficient condition for the spontaneous formation of excitons is that the corresponding binding energy, $E_{bind}$, is larger in absolute value than the band gap since then the total energy of the system can be lowered by promoting electrons to the conduction band in the absence of optical excitations \cite{bronold06,pereira22}. We computed the binding energy of an exciton in ultra-compressed hcp $^{4}$He by relying on the Wannier-Mott model (Methods) since the dielectric constant of solid helium in the TPa regime is relatively high ($\epsilon_{r} > 5$, Fig.\ref{fig:mass}) and consequently electric field screening effects are large \cite{rocca03}. 

Our excitonic binding energy results obtained with the hybrid B3LYP functional and expressed as a function of pressure are shown in Fig.\ref{fig:orbit}d. It was found that the sufficient condition for spontaneous formation of excitons, namely, $|E_{bind}| > E_{g}$, was fulfilled over a wide pressure interval of approximately $200$~GPa prior to metallization. In view of this result, we performed many-body perturbation theory calculations within the GW approximation to explicitly and more accurately determine quasiparticle excitations in ultracompressed $^{4}$He (Methods) \cite{PhysRevB.62.4927}. As it is shown in the inset of Fig.\ref{fig:orbit}d, GW calculations provided a much larger excitonic binding energy than calculated with the Wannier-Mott model and hybrid DFT functionals, namely, $\approx 0.4$~eV. Moreover, the estimated pressure interval in which excitons can spontaneously form noticeably increased up to $600$~GPa. Therefore, based in our hybrid DFT and many-body perturbation GW calculations we may conclude that on the verge of metallization hcp $^{4}$He is a bulk excitonic insulator (EI). The same conclusion was reached when considering alternative structural phases for ultra-compressed solid helium (Fig.\ref{fig:bcc}). 

\begin{figure*}[htp]
\centering
\includegraphics[width=1.00\textwidth]{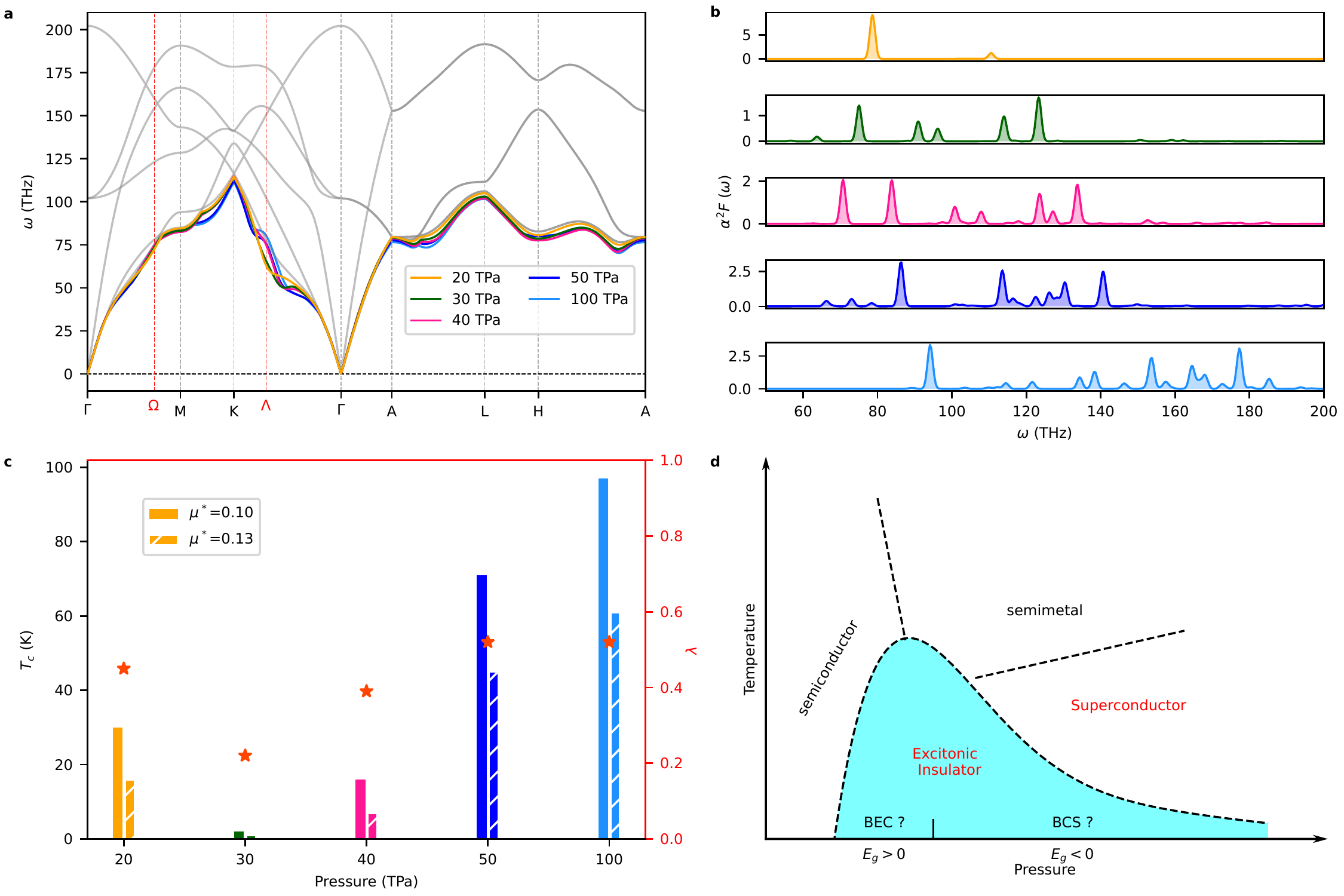}
\caption{\textbf{Electron-phonon coupling and superconducting properties of ultra-compressed hcp $^{4}$He.} \textbf{a}~Pressure-induced variation of  the acoustic phonon branches (colored lines). The phonon calculations were performed with the semi-local PBEsol functional. The grey curves were calculated for the insulating phase at $15$~TPa; the colored lines correspond to higher pressures but were re-scaled to facilitate the comparison. \textbf{b}~Eliashberg spectral function, $\alpha^2F$, of metallic hcp $^{4}$He. \textbf{c}~Superconducting properties estimated with the semi-local PBEsol functional and parameters $\mu^*=0.10$ and $\mu^*=0.13$ (Methods). The critical superconducting temperature values, $T_{c}$ (colored bars), obtained with the modified Allen-Dynes formula \cite{PhysRevB.104.054501} and the electron-phonon coupling strength parameter, $\lambda$ (red stars), are represented in the left and right ordinate axis, respectively. \textbf{d}~Qualitative sketch of the possible phase diagram of ultra-compressed solid helium based on work \cite{bronold06} and the key physical findings presented in this work.}
\label{fig:elph}
\end{figure*}

The emergence of a bulk EI state is expected to be accompanied by strong lattice distortions and instabilities due to arising electron-phonon interactions \cite{nakano18,hedayat19}. We computed the phonon spectrum of hcp $^{4}$He at different pressures using the semi-local PBE and PBEsol functionals (phonon calculations at this level of theory are feasible), as shown in Fig.\ref{fig:elph}a and Fig.\ref{fig:phonon}. Reassuringly, a distinct phonon softening appears at the reciprocal lattice point $\Lambda$ between $15$ and $20$~TPa, that is, when semi-local DFT functionals predict that solid helium becomes a metal (Fig.\ref{fig:PBEsol_band}). Interestingly, above $30$~TPa additional phonon softenings emerge along the $K$--$\Gamma$ and $M$--$K$ reciprocal space directions; we found that around this pressure the energy gap between the CBM (located at $\Gamma$) and secondary VBM (located at $\Omega$, Fig.\ref{fig:orbit}c and Fig.\ref{fig:PBEsol_band}) vanished. Thus, in addition to validating our prediction for the stabilization of a bulk EI state, these findings corroborate the strong coupling between electrons and lattice vibrations previously disclosed in ultra-compressed solid helium \cite{monserratElectronPhononCouplingMetallization2014a}. It is worth noting that quantum anharmonic effects in $^{4}$He were assessed with the stochastic self-consistent harmonic approximation (SSCHA) method~\cite{erreaAnharmonicFreeEnergies2014a,biancoSecondorderStructuralPhase2017,monacelliPressureStressTensor2018,monacelliStochasticSelfconsistentHarmonic2021}, finding that these are of little relevance in the TPa regime (Methods and Fig.\ref{fig:sscha-2}).

Besides lattice dynamics, the elastic, structural and thermodynamic properties of solid helium were also found to be influenced by the pressure-driven EI to metal phase transition (Fig.\ref{fig:cij}). In the insulating phase, the elastic constants of hcp $^{4}$He display a practically linear dependence on pressure whereas in the metallic phase they depart from this behaviour and in some cases do not even display a monotonic increase under compression (e.g., $C_{13}$). Similar effects were also observed for the bulk and shear moduli, sound velocities, Debye temperature and heat capacity (Fig.\ref{fig:cij}). Regarding the structural features, it was found that the pressure evolution of the hcp lattice parameter ratio $c/a$ drastically changes when the metallic phase is stabilized (Fig.\ref{fig:eos}). Thus, taking into account these unanticipated physical effects could have important consequences on current modelling of astrophysical bodies, in particular, of small-radius and helium-rich WDs \cite{wd1,wd2}.   

Motivated by the findings described above, we explored the superconducting properties of ultra-compressed hcp $^{4}$He. Accurate electron-phonon coupling (EPC) calculations were carried out using the techniques outlined in the Methods section, which essentially involve the Bardeen-Cooper-Schrieffer (BCS) theory of superconductivity and modified Allen-Dynes formula \cite{PhysRevB.104.054501}. Figure~\ref{fig:elph}b shows the Eliashberg spectral function, $\alpha^{2}F$, estimated at different pressures, from which the corresponding average EPC strength, $\lambda$, can be straightforwardly computed (Methods and Supplementary Information). At $20$~TPa, this function exhibits two appreciable peaks: the most prominent appearing at low frequencies stems from the lowest-energy phonon band at the wave-vector ${\bf q}= (0.25,0.43,0)$, which is very close to the reciprocal space point $\Lambda$ associated with phonon softening; the other peak emerging at higher frequencies stems from the second and third lowest-energy phonon bands at $\Gamma$. The EPC strength of these phonon modes, specially of that rendering the $\alpha^{2}F$ maximum, are extremely high (i.e., the corresponding EPC strengths, $\lambda_{q\nu}$, are of the order of $10^{1}$--$10^{2}$) due to their huge phonon linewidth and minute density of electronic states at the Fermi level (Methods, Supplementary Information and Fig.\ref{fig:PBEsol_band}). However, since the number of phonon modes that appreciably contribute to $\alpha^{2}F$ (or, equivalently, to $\lambda$) is quite reduced, the superconducting temperature estimated at $P = 20$~TPa is relatively low, namely, $T_{c} = 30$~K (Fig.\ref{fig:elph}c and Supplementary Information). 

Interestingly, upon further compression, when the energy overlap between the conduction and valence bands is enhanced (Fig.\ref{fig:orbit}c), additional peaks appear in the Eliashberg spectral function that noticeably contribute to the average EPC strength, thus raising the superconducting critical temperature. For instance, at a pressure of $50$~TPa, when multiple phonon softenings and $\alpha^{2}F$ local maxima are observed (Figs.\ref{fig:elph}a,b), we estimated a substantial superconducting critical temperature of $71$~K (Fig.\ref{fig:elph}c and Supplementary Information). Under higher compression the superconducting critical temperature steadily increases, reaching a peak value of $\approx 100$~K at the maximum pressure of $100$~TPa considered in our calculations (Fig.\ref{fig:elph}c). It is worth noting that within the pressure interval $20 \le P \le 30$~TPa both $T_c$ and $\lambda$ noticeably decrease; this transient effect is due to a dominant $P$-induced surge in the Fermi density of electronic states that drastically reduces the $\alpha^{2}F$ peaks (Fig.\ref{fig:elph}b, Methods and Table~\ref{table:tc}).

An analogous EPC strength parameter and $T_{c}$ analysis was carried out for hcp xenon (Fig.\ref{fig:Xe}) since this material is isoelectronic to solid helium and becomes metallic at experimentally accessible pressures of the order of $100$~GPa. An EI to metal phase transition similar to that disclosed in ultra-compressed $^{4}$He was also found for hcp Xe at $140$~GPa. A noticeable phonon softening appeared at a higher pressure of $190$~GPa, coinciding with the closure of a secondary band gap involving a $s$-like dominant CBM and $p$-like dominant VBM (i.e., of the same character than the primary band gap in solid helium). The EPC strength and superconducting critical temperature estimated for hcp Xe at $140$~GPa are $0.75$ and $\approx 10$~K, respectively. Thus, bulk Xe seems to be a good candidate material in which to experimentally search for analogues of some of the key theoretical findings revealed in this work for ultra-compressed hcp $^{4}$He.  

Figure~\ref{fig:elph}d shows a sketch of the possible phase diagram of solid helium at pressures and temperatures that are relevant to astrophysical studies. At sufficiently high pressures and low temperatures, a bulk EI state is stabilized. Whether in such a state the spontaneously created electron-hole bound pairs form excitonic Bose-Einstein condensates or exhibit excitonic superconductivity close to zero temperature \cite{novoselov14,bronold06}, is a matter that we cannot resolve with the DFT-based methods employed in this work (hence the questions marks in the figure). Upon further compression, hcp $^{4}$He becomes a superconductor with a critical temperature that increases under pressure (made the exception of a small pressure interval following metallization, which has been neglected in the figure). At high enough temperatures, superconductor solid helium transforms into a $p$-type semimetal. These electronic phase transitions significantly impact the structural, elastic, thermodynamic and transport properties of hcp $^{4}$He hence should be taken into consideration in advanced evolutionary models of stellar bodies like white dwarfs (WDs).

In conclusion, we have presented a comprehensive first-principles computational study of the physical properties of solid helium in the TPa regime, putting special emphasis on its electronic band-structure features. It was found that over a broad pressure range preceding metallization hcp $^{4}$He becomes a bulk excitonic insulator in which electrostatically bound electron-hole pairs can form spontaneously. This bulk excitonic insulator state could host genuine quantum many-body phenomena like high-temperature excitonic superconductivity and excitonic BEC--BCS crossover, although additional advanced studies are necessary to fully assess these hypotheses. Upon band-gap closure, solid helium transitions into a superconductor state that possesses a critical temperature of the order of $10^{1}$--$10^{2}$~K, depending on compression. This pressure-induced EI to superconductor phase transition is accompanied by several elastic and structural anomalies. Thus, our theoretical findings besides conveying great fundamental interest are also of great relevance to the physics of celestial bodies, in particular, of small-radius WDs mostly containing metallic helium. Furthermore, it is argued that some analogues of the key theoretical findings revealed here for ultra-compressed helium could be experimentally observed in solid xenon.

\section*{Methods}
\label{sec:methods}
\textbf{First-principles calculations outline.}~Density functional theory (DFT) calculations were performed with the Vienna \textit{ab initio} simulation package (VASP) \cite{kresseEfficientIterativeSchemes1996}. The projector augmented-wave (PAW) method \cite{blochlProjectorAugmentedwaveMethod1994} was employed and the $1s^2$ electrons in the He atoms were treated as valence. Different families of DFT functionals were tested among which we highlight the semi-local Perdew–Burke–Ernzerhof (PBE) \cite{perdewGeneralizedGradientApproximation1996a} and revised PBE for solids (PBEsol) \cite{perdewRestoringDensityGradientExpansion2008}, van der Waals corrected DFT-D3 \cite{grimmeConsistentAccurateInitio2010}, non-local dispersion corrected vdW-optB88 \cite{klimesChemicalAccuracyVan2009}, vdW-DF-cx\cite{berlandExchangeFunctionalThat2014}, rev-vdW-DF2 \cite{hamadaVanWaalsDensity2014}, and the hybrid HSE06\cite{krukauInfluenceExchangeScreening2006}, B3LYP \cite{stephensInitioCalculationVibrational1994} and HSEsol \cite{schimkaImprovedHybridFunctional2011}. A plane wave energy cutoff of $1500$~eV was employed along with dense Monkhorst–Pack $k$-point sampling grids of resolution $2\pi\times 0.025$~\AA~(Fig.\ref{fig:encut_test}). The energy and atomic forces in the structural relaxations were converged to within $10^{-6}$~eV and $0.002$~eV/\AA, respectively. For validation purposes, we compared our band gap results obtained with the PBE functional as implemented in the VASP code with a full-potential (linearized) augmented plane-wave method as implemented in the WIEN2k code \cite{blahaWIEN2kAPWProgram2020a} (Fig.\ref{fig:wein2k}). Phonon calculations were performed with the small displacement method and the PHONOPY code \cite{togoFirstPrinciplesPhonon2015} by employing large supercells of $4\times 4 \times 4$.

The binding energy of an exciton was estimated with the Wannier-Mott formula:
\begin{equation}
E_{bind} = - (m_u * Ry) / (m_0 \epsilon_{r}^2)~, 
\end{equation}
where $m_u= (m_e \cdot m_h) / (m_e + m_h)$. In the equation above, $m_e$ and $m_h$ are the effective mass of the electron at the bottom of the conduction band and the hole at the top of the valence band, respectively. $Ry$ represents the Rydberg constant ($= 13.6$~eV), $m_0$ the rest mass of the electron, and $\epsilon_{r}$ the dielectric constant of the system as referred to vacuum. The electron and hole effective masses were computed like the inverse of the second derivative of the conduction and valence band energies with respect to crystal momentum module, $|k|$, along the reciprocal space path $\Lambda$-$\Gamma$. The Wannier-Mott formula is a good approximation for the exciton binding energy of materials possessing high dielectric constants \cite{rocca03}, which is the case of hcp $^{4}$He in the TPa regime (Fig.\ref{fig:mass}).

The elastic tensor was determined at zero temperature by performing six finite lattice distortions and four atomic displacements of $0.01$~$\AA$ along each Cartesian direction. The adiabatic bulk modulus, $K$, and shear modulus, $G$, were obtained by computing the Voigt-Reuss-Hill averages from the elastic tensor. The longitudinal and transverse sound velocities were calculated with the formulas $v_{p}=\left[ (K+\frac{4}{3}G)/\rho \right]^{1/2}$ and $v_{s}= \left[ G/\rho \right]^{1/2}$, respectively, where $\rho$ represents the atomic density of the system. 
\\

\textbf{Crystal structure prediction analysis.}~The \textit{ab initio} random structure searching (AIRSS) package \cite{airss1,airss2} was used to perform crystal structure searches for solid $^{4}$He. The first-principles DFT code CASTEP \cite{castep} was employed to perform the underlying electronic structure calculations based on the PBE functional \cite{perdewGeneralizedGradientApproximation1996a}. The searches were performed at $100$~TPa, producing approximately $1000$~relaxed structures and considering a total of $12$ atoms in the simulation cell. The energy cutoff was set to $1000$~eV and a specially designed hard OTFG potential was employed for the calculations. Under these conditions, it was found that the hexagonal $P6_{3}/mmc$ (hcp) phase remained the ground state followed by a rhombohedral $R\overline{3}m$ phase with a higher relative energy of $0.131$~eV/atom. Other energetically competitive structures were an hexagonal $P\overline{6}m_{2}$ ($0.140$~eV/atom) and a cubic $Im\overline{3}m$ ($0.170$~eV/atom) phase.  
\\

\textbf{SSCHA calculations.}~Quantum anharmonic effects were assessed with the  stochastic self-consistent harmonic approximation (SSCHA) method~\cite{erreaAnharmonicFreeEnergies2014a,biancoSecondorderStructuralPhase2017,monacelliPressureStressTensor2018,monacelliStochasticSelfconsistentHarmonic2021}. All the SSCHA calculations were evaluated at the pressure of $35$~TPa and $0$~K, conditions at which excellent convergence of the SSCHA minimization has been verified by an extra population including $800$ supercell configurations. 
SSCHA calculations were performed with a 6$\times$6$\times$3 supercell including $216$ atoms, which yields the dynamical
matrices on a commensurate \textbf{q}-mesh of  6$\times$6$\times$3. The trial harmonic dynamical matrices used for initializing the free energy were obtained from the DFPT method as implemented in the Quantum Espresso (QE) code in the
corresponding commensurate \textbf{q}-mesh. In the self-consistent calculations of the supercells, we used the same cutoff energy as the electron-phonon coupling calculations for the primitive cell, but the \textbf{k}-mesh was reduced accordingly and was tested for convergence. In the SSCHA iterations, except the first four populations in which only internal coordinates were optimized to speed up the minimization, the free energy in other populations was minimized with respect to all degrees of freedom of the crystal structure including the internal coordinates and the lattice cell parameters. 
\\

\textbf{Many-body perturbation theory calculations.}~The excitonic binding energy was also estimated by means of highly accurate many-body perturbation theory calculations \cite{RevModPhys.74.601} performed with the Yambo code \cite{MARINI20091392}. For this, we employed the generalized gradient approximation (GGA) as parameterized by PBE together with a plane-wave basis set and norm-conserving pseudopotential. The kinetic energy cutoff for the wave functions was set to $600$~Ry. The Brillouin zone was sampled with a $64 \times 64 \times 32$ k-mesh. Many-body quasiparticle GW corrections \cite{PhysRevB.34.5390} were calculated within the single-shot G$_{0}$W$_{0}$ approximation, and the dynamic dielectric function was obtained with the plasmon-pole approximation. The exciton energies were calculated by solving the Bethe–Salpeter equation \cite{PhysRevB.62.4927} within the Tamm–Dancoff approximation \cite{PhysRev.78.382}. The static screening in the direct term was calculated within the random-phase approximation with the inclusion of local field effects. We used $2$ valence and $3$ conduction bands to solve the Bethe-Salpeter equation matrix. For the GW band-structure calculations, we sampled the Brillouin zone with a $16 \times 16 \times 8$ ${\bf k}$-point grid. A kinetic energy cutoff of $90$~Ry was used for the evaluation of the exchange part of the self-energy and of $150$~Ry for the dielectric screening matrix size. About one hundred unoccupied bands were used to build the polarizability and integrate the self-energy. The exciton energies were mapped along high symmetry paths for different pressures, as it is shown in Fig.\ref{fig:exciton} and summarized in Table~\ref{table:exciton}.
\\

\textbf{Electron-phonon coupling parameters and critical superconducting temperature.}~Electron phonon coupling (EPC) calculations were performed with the Quantum Espresso (QE) code \cite{giannozziQUANTUMESPRESSOModular2009a,giannozziAdvancedCapabilitiesMaterials2017} by using ultrasoft pseudopotentials, an energy cutoff of $200$~Ry for the kinetic energy and an energy cutoff of $2000$~Ry for the charge density (convergence tests are shown in Table~\ref{table:encut}). The equation of state of hcp $^{4}$He computed with the VASP and QE codes show very good agreement, as it is illustrated in Fig.\ref{fig:eos}. The electron–phonon matrix elements were calculated in a $16 \times 16 \times 8$ ${\bf q}$-point grid with density functional perturbation theory (DFPT) \cite{baroniPhononsRelatedCrystal2001}. We adopted a dense and shifted $k$-point mesh of $80 \times 80 \times 40$ to increase the convergence in the self-consistent calculations. For the EPC calculations, we further increased the $k$-point mesh up to $192 \times 192 \times 96$ (convergence tests are shown in Table~\ref{table:kmesh}) and to ensure $k$-point sampling convergence we employed the Methfessel–Paxton scheme with a smearing width of $0.02$~Ry. The Dirac deltas on the band energies were substituted by Gaussian functions with a broadening of $0.002$~Ry, which are necessary for the calculation of the EPC strength parameter $\lambda$ (convergence tests are shown in Fig.\ref{fig:kmesh}). Convergence tests on the ${\bf q}$-point grid sampling are also presented in Table~\ref{table:qmesh}. For further validation purposes, we also performed calculations with a hardcore pseudopotential involving a real-space cutoff of $r_{c} = 0.37~a_{0}$ \cite{PhysRevB.93.035121} and compared the results with those obtained with ultrasoft pseudopotentials, as it is shown in Fig.\ref{fig:hardcore}.

The Eliashberg spectral function, $\alpha^2F(\omega)$, accounts for the coupling between phonons and electrons in the Fermi surface like:
\begin{equation}
\alpha^2F(\omega)=\frac{1}{2\pi\hbar N(E_F)N_{q\nu}}\sum_{q\nu}\frac{\gamma_{q\nu}}{\omega_{q\nu}}\delta(\omega-\omega_{q\nu}),
\end{equation}
where $N(E_F)$ is the density of states at the Fermi level (per unit cell), $\gamma_{q\nu}$ the linewidth of the phonon mode $\nu$ at the wave vector $q$, and $N_{q\nu}$ the total number of $q\nu$ points in the sum.

The critical superconductor temperature, $T_c$, was estimated with three different formulas: the McMillan formula \cite{dynesMcMillanEquationTc1972}, $T_c^{McM}$, the Allen-Dynes formula \cite{allenTransitionTemperatureStrongcoupled1975}, $T_c^{AD}$, and the modified Allen-Dynes formula \cite{PhysRevB.104.054501}, $T_c^{mAD}$:
\begin{equation}
T_c^{McM} = \frac{\omega_{log}}{1.20}\times \exp{\left[ -\frac{1.04\left( 1 + \lambda \right)}{\lambda - \mu^{*}\left( 1 + 0.62\lambda \right)}\right]}~,
\label{eq:McM}
\end{equation}
\begin{equation}
    T_c^{AD} = f_{1}f_{2}T_c^{McM}~,
\label{eq:AD}
\end{equation}
\begin{equation}
    T_c^{mAD} = (1.0061+0.0663\lambda)T_c^{AD}~,
\label{eq:mAD}
\end{equation}
where $\mu^*$ is the Coulomb pseudopotential, for which we selected values within the widely accepted range of $0.10$--$0.13$, and the parameters $f_{1}$ and $f_{2}$ are defined like:
\begin{eqnarray}
    f_{1} = \left[1+\left(\lambda / \Lambda_{1} \right)^{3/2} \right]^{1/3} \nonumber \\
    f_{2} = 1 + \frac{\left(\bar{\omega}_{2}/\omega_{log} -1\right)\lambda^{2}}{\lambda^{2} + \Lambda_{2}^{2}}~,
\end{eqnarray}
with $\bar{\omega}_{2} = \langle \omega^{2} \rangle^{1/2}$, $\Lambda_{1} = 2.46 \left( 1 + 3.8\mu^{*} \right)$ and $\Lambda_{2} = 1.82 \left( 1 + 6.3\mu^{*}\right)\left( \bar{\omega}_{2}/\omega_{log} \right)$. %The McMillan critical superconducting temperature formula, $T_{c}^{M}$, is a particular case of $T_{c}$ in which $f_{1}=f_{2}=1$. 

Meanwhile, the logarithmic average phonon frequency, $\omega_{log}$, is defined like:
\begin{equation}
\omega_{log}=\exp{ \left[\frac{2}{\lambda}\int_{0}^{\infty}\frac{d\omega}{\omega}\alpha^2F(\omega)ln(\omega)\right]}~,
\end{equation}
and the EPC strength, $\lambda$, is proportional to the first inverse momentum of the spectral function, namely:
\begin{equation}
\lambda=2\int_{0}^{\infty}\frac{d\omega}{\omega}\alpha^2F(\omega)=\frac{1}{N_{q\nu}}\sum_{q\nu}\lambda_{q\nu}~,
\end{equation}
where
\begin{equation}
\lambda_{q\nu}=\frac{\gamma_{q\nu}}{\pi\hbar N(E_F)\omega_{q\nu} ^2}~.
\end{equation}
The $\lambda_{q\nu}$ parameter in the equation above corresponds to the EPC strength of the phonon mode at wave vector $q$ and phonon branch $\nu$. All electron phonon coupling results are listed in Table~\ref{table:tc}.
\\

\section*{Acknowledgements}
C.C. acknowledges support from the Spanish Ministry of Science, Innovation and Universities under the fellowship RYC2018-024947-I. C.L and C.C. thankfully acknowledge the computer resources at MareNostrum and the technical support provided by Barcelona Supercomputing Center (RES-FI-1-0006 and RES-FI-2022-2-0003). J.S. gratefully acknowledges the financial support from the National Key R\&D Program of China (grant nos. 2022YFA1403201), the National Natural Science Foundation of China (grant nos. 12125404, 11974162, and 11834006), and the Fundamental Research Funds for the Central Universities. Part of the calculations were carried out using supercomputers at the High Performance Computing Center of Collaborative Innovation Center of Advanced Microstructures, the high-performance supercomputing center of Nanjing University. L.J.L gratefully acknowledges the computational resources provided by the National Supercomputer Service through the United Kingdom Car-Parrinello Consortium (EP/P022561/1). I.E. and Y.-W.F. acknowledge funding from the European Research Council (ERC) under the European Union’s Horizon 2020 research and innovation program (Grant Agreement No. 802533) and the Department of Education, Universities and Research of the Eusko Jaurlaritza and the University of the Basque Country UPV/EHU (Grant No. IT1527-22). C.L and C.C. acknowledge interesting discussions and kind assistance from Raymond C. Clay III on ultra-compressed helium pseudopotentials. 
\\

%\section*{Author contributions}
%C.C. and J.B. conceived the study and planned the research. C.L. performed the DFT and EPC calculations and analysis. C.P. performed the AIRSS calculations and analysis. C.D. and J.S. performed the all-electron and many-body perturbation GW calculations and analysis. Y.W. and I.O. performed the SSCHA calculations and analysis. I.O., Y.W., C.P., B.M., L.J.C., C.D. and J.S. crucially assisted on the electron-phonon coupling and superconducting critical temperature calculations. The manuscript was written by C.L. and C.C. with substantial input from the rest of co-authors.

\section*{Additional information}
Supplementary Information is available in the online version of the paper.
\\

\section*{Competing financial interests}
The authors declare no competing financial interests.
\\

\bibliography{He.bib}

\begin{thebibliography}{48}%
\makeatletter
\providecommand \@ifxundefined [1]{%
 \@ifx{#1\undefined}
}%
\providecommand \@ifnum [1]{%
 \ifnum #1\expandafter \@firstoftwo
 \else \expandafter \@secondoftwo
 \fi
}%
\providecommand \@ifx [1]{%
 \ifx #1\expandafter \@firstoftwo
 \else \expandafter \@secondoftwo
 \fi
}%
\providecommand \natexlab [1]{#1}%
\providecommand \enquote  [1]{``#1''}%
\providecommand \bibnamefont  [1]{#1}%
\providecommand \bibfnamefont [1]{#1}%
\providecommand \citenamefont [1]{#1}%
\providecommand \href@noop [0]{\@secondoftwo}%
\providecommand \href [0]{\begingroup \@sanitize@url \@href}%
\providecommand \@href[1]{\@@startlink{#1}\@@href}%
\providecommand \@@href[1]{\endgroup#1\@@endlink}%
\providecommand \@sanitize@url [0]{\catcode `\\12\catcode `\$12\catcode
  `\&12\catcode `\#12\catcode `\^12\catcode `\_12\catcode `\%12\relax}%
\providecommand \@@startlink[1]{}%
\providecommand \@@endlink[0]{}%
\providecommand \url  [0]{\begingroup\@sanitize@url \@url }%
\providecommand \@url [1]{\endgroup\@href {#1}{\urlprefix }}%
\providecommand \urlprefix  [0]{URL }%
\providecommand \Eprint [0]{\href }%
\providecommand \doibase [0]{https://doi.org/}%
\providecommand \selectlanguage [0]{\@gobble}%
\providecommand \bibinfo  [0]{\@secondoftwo}%
\providecommand \bibfield  [0]{\@secondoftwo}%
\providecommand \translation [1]{[#1]}%
\providecommand \BibitemOpen [0]{}%
\providecommand \bibitemStop [0]{}%
\providecommand \bibitemNoStop [0]{.\EOS\space}%
\providecommand \EOS [0]{\spacefactor3000\relax}%
\providecommand \BibitemShut  [1]{\csname bibitem#1\endcsname}%
\let\auto@bib@innerbib\@empty
%</preamble>
\bibitem [{\citenamefont {Khairallah}\ and\ \citenamefont
  {Militzer}(2008)}]{khairallahFirstPrinciplesStudiesMetallization2008a}%
  \BibitemOpen
  \bibfield  {author} {\bibinfo {author} {\bibfnamefont {S.~A.}\ \bibnamefont
  {Khairallah}}\ and\ \bibinfo {author} {\bibfnamefont {B.}~\bibnamefont
  {Militzer}},\ }\bibfield  {title} {\bibinfo {title} {First-{{Principles
  Studies}} of the {{Metallization}} and the {{Equation}} of {{State}} of
  {{Solid Helium}}},\ }\href {https://doi.org/10.1103/PhysRevLett.101.106407}
  {\bibfield  {journal} {\bibinfo  {journal} {Phys. Rev. Lett.}\ }\textbf
  {\bibinfo {volume} {101}},\ \bibinfo {pages} {106407} (\bibinfo {year}
  {2008})}\BibitemShut {NoStop}%
\bibitem [{\citenamefont {Monserrat}\ \emph {et~al.}(2014)\citenamefont
  {Monserrat}, \citenamefont {Drummond}, \citenamefont {Pickard},\ and\
  \citenamefont {Needs}}]{monserratElectronPhononCouplingMetallization2014a}%
  \BibitemOpen
  \bibfield  {author} {\bibinfo {author} {\bibfnamefont {B.}~\bibnamefont
  {Monserrat}}, \bibinfo {author} {\bibfnamefont {N.~D.}\ \bibnamefont
  {Drummond}}, \bibinfo {author} {\bibfnamefont {C.~J.}\ \bibnamefont
  {Pickard}},\ and\ \bibinfo {author} {\bibfnamefont {R.~J.}\ \bibnamefont
  {Needs}},\ }\bibfield  {title} {\bibinfo {title} {Electron-{{Phonon
  Coupling}} and the {{Metallization}} of {{Solid Helium}} at {{Terapascal
  Pressures}}},\ }\href {https://doi.org/10.1103/PhysRevLett.112.055504}
  {\bibfield  {journal} {\bibinfo  {journal} {Phys. Rev. Lett.}\ }\textbf
  {\bibinfo {volume} {112}},\ \bibinfo {pages} {055504} (\bibinfo {year}
  {2014})}\BibitemShut {NoStop}%
\bibitem [{\citenamefont {Mao}\ \emph {et~al.}(2010)\citenamefont {Mao},
  \citenamefont {Shirley}, \citenamefont {Ding}, \citenamefont {Eng},
  \citenamefont {Cai}, \citenamefont {Chow}, \citenamefont {Xiao},
  \citenamefont {Shu}, \citenamefont {Hemley}, \citenamefont {Kao},\ and\
  \citenamefont {Mao}}]{maoElectronicStructureCrystalline2010}%
  \BibitemOpen
  \bibfield  {author} {\bibinfo {author} {\bibfnamefont {H.~K.}\ \bibnamefont
  {Mao}}, \bibinfo {author} {\bibfnamefont {E.~L.}\ \bibnamefont {Shirley}},
  \bibinfo {author} {\bibfnamefont {Y.}~\bibnamefont {Ding}}, \bibinfo {author}
  {\bibfnamefont {P.}~\bibnamefont {Eng}}, \bibinfo {author} {\bibfnamefont
  {Y.~Q.}\ \bibnamefont {Cai}}, \bibinfo {author} {\bibfnamefont
  {P.}~\bibnamefont {Chow}}, \bibinfo {author} {\bibfnamefont {Y.}~\bibnamefont
  {Xiao}}, \bibinfo {author} {\bibfnamefont {J.}~\bibnamefont {Shu}}, \bibinfo
  {author} {\bibfnamefont {R.~J.}\ \bibnamefont {Hemley}}, \bibinfo {author}
  {\bibfnamefont {C.}~\bibnamefont {Kao}},\ and\ \bibinfo {author}
  {\bibfnamefont {W.~L.}\ \bibnamefont {Mao}},\ }\bibfield  {title} {\bibinfo
  {title} {Electronic {Structure} of {Crystalline} {He} at {High}
  {Pressures}},\ }\href {https://doi.org/10.1103/PhysRevLett.105.186404}
  {\bibfield  {journal} {\bibinfo  {journal} {Phys. Rev. Lett.}\ }\textbf
  {\bibinfo {volume} {105}},\ \bibinfo {pages} {186404} (\bibinfo {year}
  {2010})}\BibitemShut {NoStop}%
\bibitem [{\citenamefont {Cazorla}\ and\ \citenamefont
  {Boronat}(2017)}]{cazorla17}%
  \BibitemOpen
  \bibfield  {author} {\bibinfo {author} {\bibfnamefont {C.}~\bibnamefont
  {Cazorla}}\ and\ \bibinfo {author} {\bibfnamefont {J.}~\bibnamefont
  {Boronat}},\ }\bibfield  {title} {\bibinfo {title} {Simulation and
  understanding of atomic and molecular quantum crystals},\ }\href
  {https://doi.org/10.1103/RevModPhys.89.035003} {\bibfield  {journal}
  {\bibinfo  {journal} {Rev. Mod. Phys.}\ }\textbf {\bibinfo {volume} {89}},\
  \bibinfo {pages} {035003} (\bibinfo {year} {2017})}\BibitemShut {NoStop}%
\bibitem [{\citenamefont {J{\'e}rome}\ \emph {et~al.}(1967)\citenamefont
  {J{\'e}rome}, \citenamefont {Rice},\ and\ \citenamefont
  {Kohn}}]{jeromeExcitonicInsulator1967}%
  \BibitemOpen
  \bibfield  {author} {\bibinfo {author} {\bibfnamefont {D.}~\bibnamefont
  {J{\'e}rome}}, \bibinfo {author} {\bibfnamefont {T.~M.}\ \bibnamefont
  {Rice}},\ and\ \bibinfo {author} {\bibfnamefont {W.}~\bibnamefont {Kohn}},\
  }\bibfield  {title} {\bibinfo {title} {Excitonic {{Insulator}}},\ }\href
  {https://doi.org/10.1103/PhysRev.158.462} {\bibfield  {journal} {\bibinfo
  {journal} {Phys. Rev.}\ }\textbf {\bibinfo {volume} {158}},\ \bibinfo {pages}
  {462} (\bibinfo {year} {1967})}\BibitemShut {NoStop}%
\bibitem [{\citenamefont {Ma}\ \emph {et~al.}(2021)\citenamefont {Ma},
  \citenamefont {Nguyen}, \citenamefont {Wang}, \citenamefont {Zeng},
  \citenamefont {Watanabe}, \citenamefont {Taniguchi}, \citenamefont
  {MacDonald}, \citenamefont {Mak},\ and\ \citenamefont {Shan}}]{ma21}%
  \BibitemOpen
  \bibfield  {author} {\bibinfo {author} {\bibfnamefont {L.}~\bibnamefont
  {Ma}}, \bibinfo {author} {\bibfnamefont {P.~X.}\ \bibnamefont {Nguyen}},
  \bibinfo {author} {\bibfnamefont {Z.}~\bibnamefont {Wang}}, \bibinfo {author}
  {\bibfnamefont {Y.}~\bibnamefont {Zeng}}, \bibinfo {author} {\bibfnamefont
  {K.}~\bibnamefont {Watanabe}}, \bibinfo {author} {\bibfnamefont
  {T.}~\bibnamefont {Taniguchi}}, \bibinfo {author} {\bibfnamefont {A.~H.}\
  \bibnamefont {MacDonald}}, \bibinfo {author} {\bibfnamefont {K.~F.}\
  \bibnamefont {Mak}},\ and\ \bibinfo {author} {\bibfnamefont {J.}~\bibnamefont
  {Shan}},\ }\bibfield  {title} {\bibinfo {title} {Strongly correlated
  excitonic insulator in atomic double layers},\ }\href
  {https://par.nsf.gov/biblio/10326025} {\bibfield  {journal} {\bibinfo
  {journal} {Nature}\ }\textbf {\bibinfo {volume} {598}},\ \bibinfo {pages}
  {585} (\bibinfo {year} {2021})}\BibitemShut {NoStop}%
\bibitem [{\citenamefont {Jia}\ \emph {et~al.}(2022)\citenamefont {Jia},
  \citenamefont {Wang}, \citenamefont {Chiu}, \citenamefont {Song},
  \citenamefont {Yu}, \citenamefont {Jäck}, \citenamefont {Lei}, \citenamefont
  {Klemenz}, \citenamefont {Cevallos}, \citenamefont {Onyszczak}, \citenamefont
  {Fishchenko}, \citenamefont {Liu}, \citenamefont {Farahi}, \citenamefont
  {Xie}, \citenamefont {Xu}, \citenamefont {Watanabe}, \citenamefont
  {Taniguchi}, \citenamefont {Bernevig}, \citenamefont {Cava}, \citenamefont
  {Schoop}, \citenamefont {Yazdani},\ and\ \citenamefont {Wu}}]{jia22}%
  \BibitemOpen
  \bibfield  {author} {\bibinfo {author} {\bibfnamefont {Y.}~\bibnamefont
  {Jia}}, \bibinfo {author} {\bibfnamefont {P.}~\bibnamefont {Wang}}, \bibinfo
  {author} {\bibfnamefont {C.-L.}\ \bibnamefont {Chiu}}, \bibinfo {author}
  {\bibfnamefont {Z.}~\bibnamefont {Song}}, \bibinfo {author} {\bibfnamefont
  {G.}~\bibnamefont {Yu}}, \bibinfo {author} {\bibfnamefont {B.}~\bibnamefont
  {Jäck}}, \bibinfo {author} {\bibfnamefont {S.}~\bibnamefont {Lei}}, \bibinfo
  {author} {\bibfnamefont {S.}~\bibnamefont {Klemenz}}, \bibinfo {author}
  {\bibfnamefont {F.~A.}\ \bibnamefont {Cevallos}}, \bibinfo {author}
  {\bibfnamefont {M.}~\bibnamefont {Onyszczak}}, \bibinfo {author}
  {\bibfnamefont {N.}~\bibnamefont {Fishchenko}}, \bibinfo {author}
  {\bibfnamefont {X.}~\bibnamefont {Liu}}, \bibinfo {author} {\bibfnamefont
  {G.}~\bibnamefont {Farahi}}, \bibinfo {author} {\bibfnamefont
  {F.}~\bibnamefont {Xie}}, \bibinfo {author} {\bibfnamefont {Y.}~\bibnamefont
  {Xu}}, \bibinfo {author} {\bibfnamefont {K.}~\bibnamefont {Watanabe}},
  \bibinfo {author} {\bibfnamefont {T.}~\bibnamefont {Taniguchi}}, \bibinfo
  {author} {\bibfnamefont {B.~A.}\ \bibnamefont {Bernevig}}, \bibinfo {author}
  {\bibfnamefont {R.~J.}\ \bibnamefont {Cava}}, \bibinfo {author}
  {\bibfnamefont {L.~M.}\ \bibnamefont {Schoop}}, \bibinfo {author}
  {\bibfnamefont {A.}~\bibnamefont {Yazdani}},\ and\ \bibinfo {author}
  {\bibfnamefont {S.}~\bibnamefont {Wu}},\ }\bibfield  {title} {\bibinfo
  {title} {Evidence for a monolayer excitonic insulator},\ }\href@noop {}
  {\bibfield  {journal} {\bibinfo  {journal} {Nature Physics}\ }\textbf
  {\bibinfo {volume} {18}},\ \bibinfo {pages} {87} (\bibinfo {year}
  {2022})}\BibitemShut {NoStop}%
\bibitem [{\citenamefont {Fogler}\ \emph {et~al.}(2014)\citenamefont {Fogler},
  \citenamefont {Butov},\ and\ \citenamefont {Novoselov}}]{novoselov14}%
  \BibitemOpen
  \bibfield  {author} {\bibinfo {author} {\bibfnamefont {M.~M.}\ \bibnamefont
  {Fogler}}, \bibinfo {author} {\bibfnamefont {L.~V.}\ \bibnamefont {Butov}},\
  and\ \bibinfo {author} {\bibfnamefont {K.~S.}\ \bibnamefont {Novoselov}},\
  }\bibfield  {title} {\bibinfo {title} {High-temperature superfluidity with
  indirect excitons in van der waals heterostructures},\ }\href
  {https://doi.org/10.1038/ncomms5555} {\bibfield  {journal} {\bibinfo
  {journal} {Nature communications}\ }\textbf {\bibinfo {volume} {5}},\
  \bibinfo {pages} {4555} (\bibinfo {year} {2014})}\BibitemShut {NoStop}%
\bibitem [{\citenamefont {Bronold}\ and\ \citenamefont
  {Fehske}(2006)}]{bronold06}%
  \BibitemOpen
  \bibfield  {author} {\bibinfo {author} {\bibfnamefont {F.~X.}\ \bibnamefont
  {Bronold}}\ and\ \bibinfo {author} {\bibfnamefont {H.}~\bibnamefont
  {Fehske}},\ }\bibfield  {title} {\bibinfo {title} {Possibility of an
  excitonic insulator at the semiconductor-semimetal transition},\ }\href
  {https://doi.org/10.1103/PhysRevB.74.165107} {\bibfield  {journal} {\bibinfo
  {journal} {Phys. Rev. B}\ }\textbf {\bibinfo {volume} {74}},\ \bibinfo
  {pages} {165107} (\bibinfo {year} {2006})}\BibitemShut {NoStop}%
\bibitem [{\citenamefont {{Istrate, A. G.}}\ \emph {et~al.}(2016)\citenamefont
  {{Istrate, A. G.}}, \citenamefont {{Marchant, P.}}, \citenamefont {{Tauris,
  T. M.}}, \citenamefont {{Langer, N.}}, \citenamefont {{Stancliffe, R. J.}},\
  and\ \citenamefont {{Grassitelli, L.}}}]{wd1}%
  \BibitemOpen
  \bibfield  {author} {\bibinfo {author} {\bibnamefont {{Istrate, A. G.}}},
  \bibinfo {author} {\bibnamefont {{Marchant, P.}}}, \bibinfo {author}
  {\bibnamefont {{Tauris, T. M.}}}, \bibinfo {author} {\bibnamefont {{Langer,
  N.}}}, \bibinfo {author} {\bibnamefont {{Stancliffe, R. J.}}},\ and\ \bibinfo
  {author} {\bibnamefont {{Grassitelli, L.}}},\ }\bibfield  {title} {\bibinfo
  {title} {Models of low-mass helium white dwarfs including gravitational
  settling, thermal and chemical diffusion, and rotational mixing},\ }\href
  {https://doi.org/10.1051/0004-6361/201628874} {\bibfield  {journal} {\bibinfo
   {journal} {A\&A}\ }\textbf {\bibinfo {volume} {595}},\ \bibinfo {pages}
  {A35} (\bibinfo {year} {2016})}\BibitemShut {NoStop}%
\bibitem [{\citenamefont {Cukanovaite}\ \emph {et~al.}(2018)\citenamefont
  {Cukanovaite}, \citenamefont {Tremblay}, \citenamefont {Freytag},
  \citenamefont {Ludwig},\ and\ \citenamefont {Bergeron}}]{wd2}%
  \BibitemOpen
  \bibfield  {author} {\bibinfo {author} {\bibfnamefont {E.}~\bibnamefont
  {Cukanovaite}}, \bibinfo {author} {\bibfnamefont {P.-E.}\ \bibnamefont
  {Tremblay}}, \bibinfo {author} {\bibfnamefont {B.}~\bibnamefont {Freytag}},
  \bibinfo {author} {\bibfnamefont {H.-G.}\ \bibnamefont {Ludwig}},\ and\
  \bibinfo {author} {\bibfnamefont {P.}~\bibnamefont {Bergeron}},\ }\bibfield
  {title} {\bibinfo {title} {{Pure-helium 3D model atmospheres of white
  dwarfs}},\ }\href {https://doi.org/10.1093/mnras/sty2383} {\bibfield
  {journal} {\bibinfo  {journal} {Monthly Notices of the Royal Astronomical
  Society}\ }\textbf {\bibinfo {volume} {481}},\ \bibinfo {pages} {1522}
  (\bibinfo {year} {2018})}\BibitemShut {NoStop}%
\bibitem [{\citenamefont {Pickard}\ and\ \citenamefont {Needs}(2006)}]{airss1}%
  \BibitemOpen
  \bibfield  {author} {\bibinfo {author} {\bibfnamefont {C.~J.}\ \bibnamefont
  {Pickard}}\ and\ \bibinfo {author} {\bibfnamefont {R.~J.}\ \bibnamefont
  {Needs}},\ }\bibfield  {title} {\bibinfo {title} {High-pressure phases of
  silane},\ }\href@noop {} {\bibfield  {journal} {\bibinfo  {journal} {Phys.
  Rev. Let.}\ }\textbf {\bibinfo {volume} {97}},\ \bibinfo {pages} {045504}
  (\bibinfo {year} {2006})}\BibitemShut {NoStop}%
\bibitem [{\citenamefont {Pickard}\ and\ \citenamefont {Needs}(2011)}]{airss2}%
  \BibitemOpen
  \bibfield  {author} {\bibinfo {author} {\bibfnamefont {C.~J.}\ \bibnamefont
  {Pickard}}\ and\ \bibinfo {author} {\bibfnamefont {R.~J.}\ \bibnamefont
  {Needs}},\ }\bibfield  {title} {\bibinfo {title} {\textit{Ab initio} random
  structure searching},\ }\href@noop {} {\bibfield  {journal} {\bibinfo
  {journal} {J. Phys.: Condens. Matter}\ }\textbf {\bibinfo {volume} {23}},\
  \bibinfo {pages} {053201} (\bibinfo {year} {2011})}\BibitemShut {NoStop}%
\bibitem [{\citenamefont {Cazorla}\ and\ \citenamefont
  {Boronat}(2015)}]{cazorla15}%
  \BibitemOpen
  \bibfield  {author} {\bibinfo {author} {\bibfnamefont {C.}~\bibnamefont
  {Cazorla}}\ and\ \bibinfo {author} {\bibfnamefont {J.}~\bibnamefont
  {Boronat}},\ }\bibfield  {title} {\bibinfo {title} {First-principles modeling
  of quantum nuclear effects and atomic interactions in solid $^4$he at high
  pressure},\ }\href@noop {} {\bibfield  {journal} {\bibinfo  {journal} {Phys.
  Rev. B}\ }\textbf {\bibinfo {volume} {91}},\ \bibinfo {pages} {024103}
  (\bibinfo {year} {2015})}\BibitemShut {NoStop}%
\bibitem [{\citenamefont {Pereira}(2022)}]{pereira22}%
  \BibitemOpen
  \bibfield  {author} {\bibinfo {author} {\bibfnamefont {V.~M.}\ \bibnamefont
  {Pereira}},\ }\bibfield  {title} {\bibinfo {title} {Topological excitons},\
  }\href@noop {} {\bibfield  {journal} {\bibinfo  {journal} {Nat. Phys.}\
  }\textbf {\bibinfo {volume} {18}},\ \bibinfo {pages} {6} (\bibinfo {year}
  {2022})}\BibitemShut {NoStop}%
\bibitem [{\citenamefont {La~Rocca}(2003)}]{rocca03}%
  \BibitemOpen
  \bibfield  {author} {\bibinfo {author} {\bibfnamefont {G.~C.}\ \bibnamefont
  {La~Rocca}},\ }\bibfield  {title} {\bibinfo {title} {Wannier-mott excitations
  in semiconductors},\ }\href {https://doi.org/10.1016/S1079-4050(03)31002-6}
  {\bibfield  {journal} {\bibinfo  {journal} {Thin films and Nanostructures}\
  }\textbf {\bibinfo {volume} {31}},\ \bibinfo {pages} {97} (\bibinfo {year}
  {2003})}\BibitemShut {NoStop}%
\bibitem [{\citenamefont {Rohlfing}\ and\ \citenamefont
  {Louie}(2000)}]{PhysRevB.62.4927}%
  \BibitemOpen
  \bibfield  {author} {\bibinfo {author} {\bibfnamefont {M.}~\bibnamefont
  {Rohlfing}}\ and\ \bibinfo {author} {\bibfnamefont {S.~G.}\ \bibnamefont
  {Louie}},\ }\bibfield  {title} {\bibinfo {title} {Electron-hole excitations
  and optical spectra from first principles},\ }\href
  {https://doi.org/10.1103/PhysRevB.62.4927} {\bibfield  {journal} {\bibinfo
  {journal} {Phys. Rev. B}\ }\textbf {\bibinfo {volume} {62}},\ \bibinfo
  {pages} {4927} (\bibinfo {year} {2000})}\BibitemShut {NoStop}%
\bibitem [{\citenamefont {Nakano}\ \emph {et~al.}(2018)\citenamefont {Nakano},
  \citenamefont {Hasegawa}, \citenamefont {Tamura}, \citenamefont {Katayama},
  \citenamefont {Tsutsui},\ and\ \citenamefont {Sawa}}]{nakano18}%
  \BibitemOpen
  \bibfield  {author} {\bibinfo {author} {\bibfnamefont {A.}~\bibnamefont
  {Nakano}}, \bibinfo {author} {\bibfnamefont {T.}~\bibnamefont {Hasegawa}},
  \bibinfo {author} {\bibfnamefont {S.}~\bibnamefont {Tamura}}, \bibinfo
  {author} {\bibfnamefont {N.}~\bibnamefont {Katayama}}, \bibinfo {author}
  {\bibfnamefont {S.}~\bibnamefont {Tsutsui}},\ and\ \bibinfo {author}
  {\bibfnamefont {H.}~\bibnamefont {Sawa}},\ }\bibfield  {title} {\bibinfo
  {title} {Antiferroelectric distortion with anomalous phonon softening in the
  excitonic insulator ${\mathrm{ta}}_{2}{\mathrm{nise}}_{5}$},\ }\href
  {https://doi.org/10.1103/PhysRevB.98.045139} {\bibfield  {journal} {\bibinfo
  {journal} {Phys. Rev. B}\ }\textbf {\bibinfo {volume} {98}},\ \bibinfo
  {pages} {045139} (\bibinfo {year} {2018})}\BibitemShut {NoStop}%
\bibitem [{\citenamefont {Hedayat}\ \emph {et~al.}(2019)\citenamefont
  {Hedayat}, \citenamefont {Sayers}, \citenamefont {Bugini}, \citenamefont
  {Dallera}, \citenamefont {Wolverson}, \citenamefont {Batten}, \citenamefont
  {Karbassi}, \citenamefont {Friedemann}, \citenamefont {Cerullo},
  \citenamefont {van Wezel}, \citenamefont {Clark}, \citenamefont {Carpene},\
  and\ \citenamefont {Da~Como}}]{hedayat19}%
  \BibitemOpen
  \bibfield  {author} {\bibinfo {author} {\bibfnamefont {H.}~\bibnamefont
  {Hedayat}}, \bibinfo {author} {\bibfnamefont {C.~J.}\ \bibnamefont {Sayers}},
  \bibinfo {author} {\bibfnamefont {D.}~\bibnamefont {Bugini}}, \bibinfo
  {author} {\bibfnamefont {C.}~\bibnamefont {Dallera}}, \bibinfo {author}
  {\bibfnamefont {D.}~\bibnamefont {Wolverson}}, \bibinfo {author}
  {\bibfnamefont {T.}~\bibnamefont {Batten}}, \bibinfo {author} {\bibfnamefont
  {S.}~\bibnamefont {Karbassi}}, \bibinfo {author} {\bibfnamefont
  {S.}~\bibnamefont {Friedemann}}, \bibinfo {author} {\bibfnamefont
  {G.}~\bibnamefont {Cerullo}}, \bibinfo {author} {\bibfnamefont
  {J.}~\bibnamefont {van Wezel}}, \bibinfo {author} {\bibfnamefont {S.~R.}\
  \bibnamefont {Clark}}, \bibinfo {author} {\bibfnamefont {E.}~\bibnamefont
  {Carpene}},\ and\ \bibinfo {author} {\bibfnamefont {E.}~\bibnamefont
  {Da~Como}},\ }\bibfield  {title} {\bibinfo {title} {Excitonic and lattice
  contributions to the charge density wave in
  $1t\ensuremath{-}\mathrm{TiS}{\mathrm{e}}_{2}$ revealed by a phonon
  bottleneck},\ }\href {https://doi.org/10.1103/PhysRevResearch.1.023029}
  {\bibfield  {journal} {\bibinfo  {journal} {Phys. Rev. Research}\ }\textbf
  {\bibinfo {volume} {1}},\ \bibinfo {pages} {023029} (\bibinfo {year}
  {2019})}\BibitemShut {NoStop}%
\bibitem [{\citenamefont {Errea}\ \emph {et~al.}(2014)\citenamefont {Errea},
  \citenamefont {Calandra},\ and\ \citenamefont
  {Mauri}}]{erreaAnharmonicFreeEnergies2014a}%
  \BibitemOpen
  \bibfield  {author} {\bibinfo {author} {\bibfnamefont {I.}~\bibnamefont
  {Errea}}, \bibinfo {author} {\bibfnamefont {M.}~\bibnamefont {Calandra}},\
  and\ \bibinfo {author} {\bibfnamefont {F.}~\bibnamefont {Mauri}},\ }\bibfield
   {title} {\bibinfo {title} {Anharmonic free energies and phonon dispersions
  from the stochastic self-consistent harmonic approximation: {{Application}}
  to platinum and palladium hydrides},\ }\href
  {https://doi.org/10.1103/PhysRevB.89.064302} {\bibfield  {journal} {\bibinfo
  {journal} {Phys. Rev. B}\ }\textbf {\bibinfo {volume} {89}},\ \bibinfo
  {pages} {064302} (\bibinfo {year} {2014})}\BibitemShut {NoStop}%
\bibitem [{\citenamefont {Bianco}\ \emph {et~al.}(2017)\citenamefont {Bianco},
  \citenamefont {Errea}, \citenamefont {Paulatto}, \citenamefont {Calandra},\
  and\ \citenamefont {Mauri}}]{biancoSecondorderStructuralPhase2017}%
  \BibitemOpen
  \bibfield  {author} {\bibinfo {author} {\bibfnamefont {R.}~\bibnamefont
  {Bianco}}, \bibinfo {author} {\bibfnamefont {I.}~\bibnamefont {Errea}},
  \bibinfo {author} {\bibfnamefont {L.}~\bibnamefont {Paulatto}}, \bibinfo
  {author} {\bibfnamefont {M.}~\bibnamefont {Calandra}},\ and\ \bibinfo
  {author} {\bibfnamefont {F.}~\bibnamefont {Mauri}},\ }\bibfield  {title}
  {\bibinfo {title} {Second-order structural phase transitions, free energy
  curvature, and temperature-dependent anharmonic phonons in the
  self-consistent harmonic approximation: {{Theory}} and stochastic
  implementation},\ }\href {https://doi.org/10.1103/PhysRevB.96.014111}
  {\bibfield  {journal} {\bibinfo  {journal} {Phys. Rev. B}\ }\textbf {\bibinfo
  {volume} {96}},\ \bibinfo {pages} {014111} (\bibinfo {year}
  {2017})}\BibitemShut {NoStop}%
\bibitem [{\citenamefont {Monacelli}\ \emph {et~al.}(2018)\citenamefont
  {Monacelli}, \citenamefont {Errea}, \citenamefont {Calandra},\ and\
  \citenamefont {Mauri}}]{monacelliPressureStressTensor2018}%
  \BibitemOpen
  \bibfield  {author} {\bibinfo {author} {\bibfnamefont {L.}~\bibnamefont
  {Monacelli}}, \bibinfo {author} {\bibfnamefont {I.}~\bibnamefont {Errea}},
  \bibinfo {author} {\bibfnamefont {M.}~\bibnamefont {Calandra}},\ and\
  \bibinfo {author} {\bibfnamefont {F.}~\bibnamefont {Mauri}},\ }\bibfield
  {title} {\bibinfo {title} {Pressure and stress tensor of complex anharmonic
  crystals within the stochastic self-consistent harmonic approximation},\
  }\href {https://doi.org/10.1103/PhysRevB.98.024106} {\bibfield  {journal}
  {\bibinfo  {journal} {Phys. Rev. B}\ }\textbf {\bibinfo {volume} {98}},\
  \bibinfo {pages} {024106} (\bibinfo {year} {2018})}\BibitemShut {NoStop}%
\bibitem [{\citenamefont {Monacelli}\ \emph {et~al.}(2021)\citenamefont
  {Monacelli}, \citenamefont {Bianco}, \citenamefont {Cherubini}, \citenamefont
  {Calandra}, \citenamefont {Errea},\ and\ \citenamefont
  {Mauri}}]{monacelliStochasticSelfconsistentHarmonic2021}%
  \BibitemOpen
  \bibfield  {author} {\bibinfo {author} {\bibfnamefont {L.}~\bibnamefont
  {Monacelli}}, \bibinfo {author} {\bibfnamefont {R.}~\bibnamefont {Bianco}},
  \bibinfo {author} {\bibfnamefont {M.}~\bibnamefont {Cherubini}}, \bibinfo
  {author} {\bibfnamefont {M.}~\bibnamefont {Calandra}}, \bibinfo {author}
  {\bibfnamefont {I.}~\bibnamefont {Errea}},\ and\ \bibinfo {author}
  {\bibfnamefont {F.}~\bibnamefont {Mauri}},\ }\bibfield  {title} {\bibinfo
  {title} {The stochastic self-consistent harmonic approximation: Calculating
  vibrational properties of materials with full quantum and anharmonic
  effects},\ }\href {https://doi.org/10.1088/1361-648X/ac066b} {\bibfield
  {journal} {\bibinfo  {journal} {J. Phys.: Condens. Matter}\ }\textbf
  {\bibinfo {volume} {33}},\ \bibinfo {pages} {363001} (\bibinfo {year}
  {2021})}\BibitemShut {NoStop}%
\bibitem [{\citenamefont {Shipley}\ \emph {et~al.}(2021)\citenamefont
  {Shipley}, \citenamefont {Hutcheon}, \citenamefont {Needs},\ and\
  \citenamefont {Pickard}}]{PhysRevB.104.054501}%
  \BibitemOpen
  \bibfield  {author} {\bibinfo {author} {\bibfnamefont {A.~M.}\ \bibnamefont
  {Shipley}}, \bibinfo {author} {\bibfnamefont {M.~J.}\ \bibnamefont
  {Hutcheon}}, \bibinfo {author} {\bibfnamefont {R.~J.}\ \bibnamefont
  {Needs}},\ and\ \bibinfo {author} {\bibfnamefont {C.~J.}\ \bibnamefont
  {Pickard}},\ }\bibfield  {title} {\bibinfo {title} {High-throughput discovery
  of high-temperature conventional superconductors},\ }\href
  {https://doi.org/10.1103/PhysRevB.104.054501} {\bibfield  {journal} {\bibinfo
   {journal} {Phys. Rev. B}\ }\textbf {\bibinfo {volume} {104}},\ \bibinfo
  {pages} {054501} (\bibinfo {year} {2021})}\BibitemShut {NoStop}%
\bibitem [{\citenamefont {Kresse}\ and\ \citenamefont
  {Furthm{\"u}ller}(1996)}]{kresseEfficientIterativeSchemes1996}%
  \BibitemOpen
  \bibfield  {author} {\bibinfo {author} {\bibfnamefont {G.}~\bibnamefont
  {Kresse}}\ and\ \bibinfo {author} {\bibfnamefont {J.}~\bibnamefont
  {Furthm{\"u}ller}},\ }\bibfield  {title} {\bibinfo {title} {Efficient
  iterative schemes for ab initio total-energy calculations using a plane-wave
  basis set},\ }\href {https://doi.org/10.1103/PhysRevB.54.11169} {\bibfield
  {journal} {\bibinfo  {journal} {Phys. Rev. B}\ }\textbf {\bibinfo {volume}
  {54}},\ \bibinfo {pages} {11169} (\bibinfo {year} {1996})}\BibitemShut
  {NoStop}%
\bibitem [{\citenamefont
  {Bl{\"o}chl}(1994)}]{blochlProjectorAugmentedwaveMethod1994}%
  \BibitemOpen
  \bibfield  {author} {\bibinfo {author} {\bibfnamefont {P.~E.}\ \bibnamefont
  {Bl{\"o}chl}},\ }\bibfield  {title} {\bibinfo {title} {Projector
  augmented-wave method},\ }\href {https://doi.org/10.1103/PhysRevB.50.17953}
  {\bibfield  {journal} {\bibinfo  {journal} {Phys. Rev. B}\ }\textbf {\bibinfo
  {volume} {50}},\ \bibinfo {pages} {17953} (\bibinfo {year}
  {1994})}\BibitemShut {NoStop}%
\bibitem [{\citenamefont {Perdew}\ \emph {et~al.}(1996)\citenamefont {Perdew},
  \citenamefont {Burke},\ and\ \citenamefont
  {Ernzerhof}}]{perdewGeneralizedGradientApproximation1996a}%
  \BibitemOpen
  \bibfield  {author} {\bibinfo {author} {\bibfnamefont {J.~P.}\ \bibnamefont
  {Perdew}}, \bibinfo {author} {\bibfnamefont {K.}~\bibnamefont {Burke}},\ and\
  \bibinfo {author} {\bibfnamefont {M.}~\bibnamefont {Ernzerhof}},\ }\bibfield
  {title} {\bibinfo {title} {Generalized {{Gradient Approximation Made
  Simple}}},\ }\href {https://doi.org/10.1103/PhysRevLett.77.3865} {\bibfield
  {journal} {\bibinfo  {journal} {Phys. Rev. Lett.}\ }\textbf {\bibinfo
  {volume} {77}},\ \bibinfo {pages} {3865} (\bibinfo {year}
  {1996})}\BibitemShut {NoStop}%
\bibitem [{\citenamefont {Perdew}\ \emph {et~al.}(2008)\citenamefont {Perdew},
  \citenamefont {Ruzsinszky}, \citenamefont {Csonka}, \citenamefont {Vydrov},
  \citenamefont {Scuseria}, \citenamefont {Constantin}, \citenamefont {Zhou},\
  and\ \citenamefont {Burke}}]{perdewRestoringDensityGradientExpansion2008}%
  \BibitemOpen
  \bibfield  {author} {\bibinfo {author} {\bibfnamefont {J.~P.}\ \bibnamefont
  {Perdew}}, \bibinfo {author} {\bibfnamefont {A.}~\bibnamefont {Ruzsinszky}},
  \bibinfo {author} {\bibfnamefont {G.~I.}\ \bibnamefont {Csonka}}, \bibinfo
  {author} {\bibfnamefont {O.~A.}\ \bibnamefont {Vydrov}}, \bibinfo {author}
  {\bibfnamefont {G.~E.}\ \bibnamefont {Scuseria}}, \bibinfo {author}
  {\bibfnamefont {L.~A.}\ \bibnamefont {Constantin}}, \bibinfo {author}
  {\bibfnamefont {X.}~\bibnamefont {Zhou}},\ and\ \bibinfo {author}
  {\bibfnamefont {K.}~\bibnamefont {Burke}},\ }\bibfield  {title} {\bibinfo
  {title} {Restoring the {{Density-Gradient Expansion}} for {{Exchange}} in
  {{Solids}} and {{Surfaces}}},\ }\href
  {https://doi.org/10.1103/PhysRevLett.100.136406} {\bibfield  {journal}
  {\bibinfo  {journal} {Phys. Rev. Lett.}\ }\textbf {\bibinfo {volume} {100}},\
  \bibinfo {pages} {136406} (\bibinfo {year} {2008})}\BibitemShut {NoStop}%
\bibitem [{\citenamefont {Grimme}\ \emph {et~al.}(2010)\citenamefont {Grimme},
  \citenamefont {Antony}, \citenamefont {Ehrlich},\ and\ \citenamefont
  {Krieg}}]{grimmeConsistentAccurateInitio2010}%
  \BibitemOpen
  \bibfield  {author} {\bibinfo {author} {\bibfnamefont {S.}~\bibnamefont
  {Grimme}}, \bibinfo {author} {\bibfnamefont {J.}~\bibnamefont {Antony}},
  \bibinfo {author} {\bibfnamefont {S.}~\bibnamefont {Ehrlich}},\ and\ \bibinfo
  {author} {\bibfnamefont {H.}~\bibnamefont {Krieg}},\ }\bibfield  {title}
  {\bibinfo {title} {A consistent and accurate ab initio parametrization of
  density functional dispersion correction ({{DFT-D}}) for the 94 elements
  {{H-Pu}}},\ }\href {https://doi.org/10.1063/1.3382344} {\bibfield  {journal}
  {\bibinfo  {journal} {The Journal of Chemical Physics}\ }\textbf {\bibinfo
  {volume} {132}},\ \bibinfo {pages} {154104} (\bibinfo {year}
  {2010})}\BibitemShut {NoStop}%
\bibitem [{\citenamefont {Klime{\v s}}\ \emph {et~al.}(2009)\citenamefont
  {Klime{\v s}}, \citenamefont {Bowler},\ and\ \citenamefont
  {Michaelides}}]{klimesChemicalAccuracyVan2009}%
  \BibitemOpen
  \bibfield  {author} {\bibinfo {author} {\bibfnamefont {J.}~\bibnamefont
  {Klime{\v s}}}, \bibinfo {author} {\bibfnamefont {D.~R.}\ \bibnamefont
  {Bowler}},\ and\ \bibinfo {author} {\bibfnamefont {A.}~\bibnamefont
  {Michaelides}},\ }\bibfield  {title} {\bibinfo {title} {Chemical accuracy for
  the van der {{Waals}} density functional},\ }\href
  {https://doi.org/10.1088/0953-8984/22/2/022201} {\bibfield  {journal}
  {\bibinfo  {journal} {J. Phys.: Condens. Matter}\ }\textbf {\bibinfo {volume}
  {22}},\ \bibinfo {pages} {022201} (\bibinfo {year} {2009})}\BibitemShut
  {NoStop}%
\bibitem [{\citenamefont {Berland}\ and\ \citenamefont
  {Hyldgaard}(2014)}]{berlandExchangeFunctionalThat2014}%
  \BibitemOpen
  \bibfield  {author} {\bibinfo {author} {\bibfnamefont {K.}~\bibnamefont
  {Berland}}\ and\ \bibinfo {author} {\bibfnamefont {P.}~\bibnamefont
  {Hyldgaard}},\ }\bibfield  {title} {\bibinfo {title} {Exchange functional
  that tests the robustness of the plasmon description of the van der {{Waals}}
  density functional},\ }\href {https://doi.org/10.1103/PhysRevB.89.035412}
  {\bibfield  {journal} {\bibinfo  {journal} {Phys. Rev. B}\ }\textbf {\bibinfo
  {volume} {89}},\ \bibinfo {pages} {035412} (\bibinfo {year}
  {2014})}\BibitemShut {NoStop}%
\bibitem [{\citenamefont {Hamada}(2014)}]{hamadaVanWaalsDensity2014}%
  \BibitemOpen
  \bibfield  {author} {\bibinfo {author} {\bibfnamefont {I.}~\bibnamefont
  {Hamada}},\ }\bibfield  {title} {\bibinfo {title} {Van der {{Waals}} density
  functional made accurate},\ }\href
  {https://doi.org/10.1103/PhysRevB.89.121103} {\bibfield  {journal} {\bibinfo
  {journal} {Phys. Rev. B}\ }\textbf {\bibinfo {volume} {89}},\ \bibinfo
  {pages} {121103} (\bibinfo {year} {2014})}\BibitemShut {NoStop}%
\bibitem [{\citenamefont {Krukau}\ \emph {et~al.}(2006)\citenamefont {Krukau},
  \citenamefont {Vydrov}, \citenamefont {Izmaylov},\ and\ \citenamefont
  {Scuseria}}]{krukauInfluenceExchangeScreening2006}%
  \BibitemOpen
  \bibfield  {author} {\bibinfo {author} {\bibfnamefont {A.~V.}\ \bibnamefont
  {Krukau}}, \bibinfo {author} {\bibfnamefont {O.~A.}\ \bibnamefont {Vydrov}},
  \bibinfo {author} {\bibfnamefont {A.~F.}\ \bibnamefont {Izmaylov}},\ and\
  \bibinfo {author} {\bibfnamefont {G.~E.}\ \bibnamefont {Scuseria}},\
  }\bibfield  {title} {\bibinfo {title} {Influence of the exchange screening
  parameter on the performance of screened hybrid functionals},\ }\href
  {https://doi.org/10.1063/1.2404663} {\bibfield  {journal} {\bibinfo
  {journal} {J. Chem. Phys.}\ }\textbf {\bibinfo {volume} {125}},\ \bibinfo
  {pages} {224106} (\bibinfo {year} {2006})}\BibitemShut {NoStop}%
\bibitem [{\citenamefont {Stephens}\ \emph {et~al.}(1994)\citenamefont
  {Stephens}, \citenamefont {Devlin}, \citenamefont {Chabalowski},\ and\
  \citenamefont {Frisch}}]{stephensInitioCalculationVibrational1994}%
  \BibitemOpen
  \bibfield  {author} {\bibinfo {author} {\bibfnamefont {P.~J.}\ \bibnamefont
  {Stephens}}, \bibinfo {author} {\bibfnamefont {F.~J.}\ \bibnamefont
  {Devlin}}, \bibinfo {author} {\bibfnamefont {C.~F.}\ \bibnamefont
  {Chabalowski}},\ and\ \bibinfo {author} {\bibfnamefont {M.~J.}\ \bibnamefont
  {Frisch}},\ }\bibfield  {title} {\bibinfo {title} {Ab {{Initio Calculation}}
  of {{Vibrational Absorption}} and {{Circular Dichroism Spectra Using Density
  Functional Force Fields}}},\ }\href {https://doi.org/10.1021/j100096a001}
  {\bibfield  {journal} {\bibinfo  {journal} {J. Phys. Chem.}\ }\textbf
  {\bibinfo {volume} {98}},\ \bibinfo {pages} {11623} (\bibinfo {year}
  {1994})}\BibitemShut {NoStop}%
\bibitem [{\citenamefont {Schimka}\ \emph {et~al.}(2011)\citenamefont
  {Schimka}, \citenamefont {Harl},\ and\ \citenamefont
  {Kresse}}]{schimkaImprovedHybridFunctional2011}%
  \BibitemOpen
  \bibfield  {author} {\bibinfo {author} {\bibfnamefont {L.}~\bibnamefont
  {Schimka}}, \bibinfo {author} {\bibfnamefont {J.}~\bibnamefont {Harl}},\ and\
  \bibinfo {author} {\bibfnamefont {G.}~\bibnamefont {Kresse}},\ }\bibfield
  {title} {\bibinfo {title} {Improved hybrid functional for solids: {{The
  HSEsol}} functional},\ }\href {https://doi.org/10.1063/1.3524336} {\bibfield
  {journal} {\bibinfo  {journal} {J. Chem. Phys.}\ }\textbf {\bibinfo {volume}
  {134}},\ \bibinfo {pages} {024116} (\bibinfo {year} {2011})}\BibitemShut
  {NoStop}%
\bibitem [{\citenamefont {Blaha}\ \emph {et~al.}(2020)\citenamefont {Blaha},
  \citenamefont {Schwarz}, \citenamefont {Tran}, \citenamefont {Laskowski},
  \citenamefont {Madsen},\ and\ \citenamefont
  {Marks}}]{blahaWIEN2kAPWProgram2020a}%
  \BibitemOpen
  \bibfield  {author} {\bibinfo {author} {\bibfnamefont {P.}~\bibnamefont
  {Blaha}}, \bibinfo {author} {\bibfnamefont {K.}~\bibnamefont {Schwarz}},
  \bibinfo {author} {\bibfnamefont {F.}~\bibnamefont {Tran}}, \bibinfo {author}
  {\bibfnamefont {R.}~\bibnamefont {Laskowski}}, \bibinfo {author}
  {\bibfnamefont {G.~K.~H.}\ \bibnamefont {Madsen}},\ and\ \bibinfo {author}
  {\bibfnamefont {L.~D.}\ \bibnamefont {Marks}},\ }\bibfield  {title} {\bibinfo
  {title} {{{WIEN2k}}: {{An APW}}+lo program for calculating the properties of
  solids},\ }\href {https://doi.org/10.1063/1.5143061} {\bibfield  {journal}
  {\bibinfo  {journal} {J. Chem. Phys.}\ }\textbf {\bibinfo {volume} {152}},\
  \bibinfo {pages} {074101} (\bibinfo {year} {2020})}\BibitemShut {NoStop}%
\bibitem [{\citenamefont {Togo}\ and\ \citenamefont
  {Tanaka}(2015)}]{togoFirstPrinciplesPhonon2015}%
  \BibitemOpen
  \bibfield  {author} {\bibinfo {author} {\bibfnamefont {A.}~\bibnamefont
  {Togo}}\ and\ \bibinfo {author} {\bibfnamefont {I.}~\bibnamefont {Tanaka}},\
  }\bibfield  {title} {\bibinfo {title} {First principles phonon calculations
  in materials science},\ }\href
  {https://doi.org/10.1016/j.scriptamat.2015.07.021} {\bibfield  {journal}
  {\bibinfo  {journal} {Scripta Materialia}\ }\textbf {\bibinfo {volume}
  {108}},\ \bibinfo {pages} {1} (\bibinfo {year} {2015})}\BibitemShut {NoStop}%
\bibitem [{\citenamefont {Clark}\ \emph {et~al.}(2005)\citenamefont {Clark},
  \citenamefont {Segall}, \citenamefont {Pickard}, \citenamefont {Hasnip},
  \citenamefont {Probert}, \citenamefont {Refson},\ and\ \citenamefont
  {Payne}}]{castep}%
  \BibitemOpen
  \bibfield  {author} {\bibinfo {author} {\bibfnamefont {S.~J.}\ \bibnamefont
  {Clark}}, \bibinfo {author} {\bibfnamefont {M.~D.}\ \bibnamefont {Segall}},
  \bibinfo {author} {\bibfnamefont {C.~J.}\ \bibnamefont {Pickard}}, \bibinfo
  {author} {\bibfnamefont {P.~J.}\ \bibnamefont {Hasnip}}, \bibinfo {author}
  {\bibfnamefont {M.~I.~J.}\ \bibnamefont {Probert}}, \bibinfo {author}
  {\bibfnamefont {K.}~\bibnamefont {Refson}},\ and\ \bibinfo {author}
  {\bibfnamefont {M.~C.}\ \bibnamefont {Payne}},\ }\bibfield  {title} {\bibinfo
  {title} {First principles methods using castep},\ }\href@noop {} {\bibfield
  {journal} {\bibinfo  {journal} {Z. Kristallogr. - Cryst. Mater.}\ }\textbf
  {\bibinfo {volume} {220}},\ \bibinfo {pages} {567} (\bibinfo {year}
  {2005})}\BibitemShut {NoStop}%
\bibitem [{\citenamefont {Onida}\ \emph {et~al.}(2002)\citenamefont {Onida},
  \citenamefont {Reining},\ and\ \citenamefont {Rubio}}]{RevModPhys.74.601}%
  \BibitemOpen
  \bibfield  {author} {\bibinfo {author} {\bibfnamefont {G.}~\bibnamefont
  {Onida}}, \bibinfo {author} {\bibfnamefont {L.}~\bibnamefont {Reining}},\
  and\ \bibinfo {author} {\bibfnamefont {A.}~\bibnamefont {Rubio}},\ }\bibfield
   {title} {\bibinfo {title} {Electronic excitations: density-functional versus
  many-body green's-function approaches},\ }\href
  {https://doi.org/10.1103/RevModPhys.74.601} {\bibfield  {journal} {\bibinfo
  {journal} {Rev. Mod. Phys.}\ }\textbf {\bibinfo {volume} {74}},\ \bibinfo
  {pages} {601} (\bibinfo {year} {2002})}\BibitemShut {NoStop}%
\bibitem [{\citenamefont {Marini}\ \emph {et~al.}(2009)\citenamefont {Marini},
  \citenamefont {Hogan}, \citenamefont {Grüning},\ and\ \citenamefont
  {Varsano}}]{MARINI20091392}%
  \BibitemOpen
  \bibfield  {author} {\bibinfo {author} {\bibfnamefont {A.}~\bibnamefont
  {Marini}}, \bibinfo {author} {\bibfnamefont {C.}~\bibnamefont {Hogan}},
  \bibinfo {author} {\bibfnamefont {M.}~\bibnamefont {Grüning}},\ and\
  \bibinfo {author} {\bibfnamefont {D.}~\bibnamefont {Varsano}},\ }\bibfield
  {title} {\bibinfo {title} {yambo: An ab initio tool for excited state
  calculations},\ }\href
  {https://doi.org/https://doi.org/10.1016/j.cpc.2009.02.003} {\bibfield
  {journal} {\bibinfo  {journal} {Computer Physics Communications}\ }\textbf
  {\bibinfo {volume} {180}},\ \bibinfo {pages} {1392} (\bibinfo {year}
  {2009})}\BibitemShut {NoStop}%
\bibitem [{\citenamefont {Hybertsen}\ and\ \citenamefont
  {Louie}(1986)}]{PhysRevB.34.5390}%
  \BibitemOpen
  \bibfield  {author} {\bibinfo {author} {\bibfnamefont {M.~S.}\ \bibnamefont
  {Hybertsen}}\ and\ \bibinfo {author} {\bibfnamefont {S.~G.}\ \bibnamefont
  {Louie}},\ }\bibfield  {title} {\bibinfo {title} {Electron correlation in
  semiconductors and insulators: Band gaps and quasiparticle energies},\ }\href
  {https://doi.org/10.1103/PhysRevB.34.5390} {\bibfield  {journal} {\bibinfo
  {journal} {Phys. Rev. B}\ }\textbf {\bibinfo {volume} {34}},\ \bibinfo
  {pages} {5390} (\bibinfo {year} {1986})}\BibitemShut {NoStop}%
\bibitem [{\citenamefont {Dancoff}(1950)}]{PhysRev.78.382}%
  \BibitemOpen
  \bibfield  {author} {\bibinfo {author} {\bibfnamefont {S.~M.}\ \bibnamefont
  {Dancoff}},\ }\bibfield  {title} {\bibinfo {title} {Non-adiabatic meson
  theory of nuclear forces},\ }\href {https://doi.org/10.1103/PhysRev.78.382}
  {\bibfield  {journal} {\bibinfo  {journal} {Phys. Rev.}\ }\textbf {\bibinfo
  {volume} {78}},\ \bibinfo {pages} {382} (\bibinfo {year} {1950})}\BibitemShut
  {NoStop}%
\bibitem [{\citenamefont {Giannozzi}\ \emph {et~al.}(2009)\citenamefont
  {Giannozzi}, \citenamefont {Baroni}, \citenamefont {Bonini}, \citenamefont
  {Calandra}, \citenamefont {Car}, \citenamefont {Cavazzoni}, \citenamefont
  {Ceresoli}, \citenamefont {Chiarotti}, \citenamefont {Cococcioni},
  \citenamefont {Dabo}, \citenamefont {Corso}, \citenamefont {de~Gironcoli},
  \citenamefont {Fabris}, \citenamefont {Fratesi}, \citenamefont {Gebauer},
  \citenamefont {Gerstmann}, \citenamefont {Gougoussis}, \citenamefont
  {Kokalj}, \citenamefont {Lazzeri}, \citenamefont {{Martin-Samos}},
  \citenamefont {Marzari}, \citenamefont {Mauri}, \citenamefont {Mazzarello},
  \citenamefont {Paolini}, \citenamefont {Pasquarello}, \citenamefont
  {Paulatto}, \citenamefont {Sbraccia}, \citenamefont {Scandolo}, \citenamefont
  {Sclauzero}, \citenamefont {Seitsonen}, \citenamefont {Smogunov},
  \citenamefont {Umari},\ and\ \citenamefont
  {Wentzcovitch}}]{giannozziQUANTUMESPRESSOModular2009a}%
  \BibitemOpen
  \bibfield  {author} {\bibinfo {author} {\bibfnamefont {P.}~\bibnamefont
  {Giannozzi}}, \bibinfo {author} {\bibfnamefont {S.}~\bibnamefont {Baroni}},
  \bibinfo {author} {\bibfnamefont {N.}~\bibnamefont {Bonini}}, \bibinfo
  {author} {\bibfnamefont {M.}~\bibnamefont {Calandra}}, \bibinfo {author}
  {\bibfnamefont {R.}~\bibnamefont {Car}}, \bibinfo {author} {\bibfnamefont
  {C.}~\bibnamefont {Cavazzoni}}, \bibinfo {author} {\bibfnamefont
  {D.}~\bibnamefont {Ceresoli}}, \bibinfo {author} {\bibfnamefont {G.~L.}\
  \bibnamefont {Chiarotti}}, \bibinfo {author} {\bibfnamefont {M.}~\bibnamefont
  {Cococcioni}}, \bibinfo {author} {\bibfnamefont {I.}~\bibnamefont {Dabo}},
  \bibinfo {author} {\bibfnamefont {A.~D.}\ \bibnamefont {Corso}}, \bibinfo
  {author} {\bibfnamefont {S.}~\bibnamefont {de~Gironcoli}}, \bibinfo {author}
  {\bibfnamefont {S.}~\bibnamefont {Fabris}}, \bibinfo {author} {\bibfnamefont
  {G.}~\bibnamefont {Fratesi}}, \bibinfo {author} {\bibfnamefont
  {R.}~\bibnamefont {Gebauer}}, \bibinfo {author} {\bibfnamefont
  {U.}~\bibnamefont {Gerstmann}}, \bibinfo {author} {\bibfnamefont
  {C.}~\bibnamefont {Gougoussis}}, \bibinfo {author} {\bibfnamefont
  {A.}~\bibnamefont {Kokalj}}, \bibinfo {author} {\bibfnamefont
  {M.}~\bibnamefont {Lazzeri}}, \bibinfo {author} {\bibfnamefont
  {L.}~\bibnamefont {{Martin-Samos}}}, \bibinfo {author} {\bibfnamefont
  {N.}~\bibnamefont {Marzari}}, \bibinfo {author} {\bibfnamefont
  {F.}~\bibnamefont {Mauri}}, \bibinfo {author} {\bibfnamefont
  {R.}~\bibnamefont {Mazzarello}}, \bibinfo {author} {\bibfnamefont
  {S.}~\bibnamefont {Paolini}}, \bibinfo {author} {\bibfnamefont
  {A.}~\bibnamefont {Pasquarello}}, \bibinfo {author} {\bibfnamefont
  {L.}~\bibnamefont {Paulatto}}, \bibinfo {author} {\bibfnamefont
  {C.}~\bibnamefont {Sbraccia}}, \bibinfo {author} {\bibfnamefont
  {S.}~\bibnamefont {Scandolo}}, \bibinfo {author} {\bibfnamefont
  {G.}~\bibnamefont {Sclauzero}}, \bibinfo {author} {\bibfnamefont {A.~P.}\
  \bibnamefont {Seitsonen}}, \bibinfo {author} {\bibfnamefont {A.}~\bibnamefont
  {Smogunov}}, \bibinfo {author} {\bibfnamefont {P.}~\bibnamefont {Umari}},\
  and\ \bibinfo {author} {\bibfnamefont {R.~M.}\ \bibnamefont {Wentzcovitch}},\
  }\bibfield  {title} {\bibinfo {title} {{{QUANTUM ESPRESSO}}: A modular and
  open-source software project for quantum simulations of materials},\ }\href
  {https://doi.org/10.1088/0953-8984/21/39/395502} {\bibfield  {journal}
  {\bibinfo  {journal} {J. Phys.: Condens. Matter}\ }\textbf {\bibinfo {volume}
  {21}},\ \bibinfo {pages} {395502} (\bibinfo {year} {2009})}\BibitemShut
  {NoStop}%
\bibitem [{\citenamefont {Giannozzi}\ \emph {et~al.}(2017)\citenamefont
  {Giannozzi}, \citenamefont {Andreussi}, \citenamefont {Brumme}, \citenamefont
  {Bunau}, \citenamefont {Nardelli}, \citenamefont {Calandra}, \citenamefont
  {Car}, \citenamefont {Cavazzoni}, \citenamefont {Ceresoli}, \citenamefont
  {Cococcioni}, \citenamefont {Colonna}, \citenamefont {Carnimeo},
  \citenamefont {Corso}, \citenamefont {de~Gironcoli}, \citenamefont {Delugas},
  \citenamefont {DiStasio}, \citenamefont {Ferretti}, \citenamefont {Floris},
  \citenamefont {Fratesi}, \citenamefont {Fugallo}, \citenamefont {Gebauer},
  \citenamefont {Gerstmann}, \citenamefont {Giustino}, \citenamefont {Gorni},
  \citenamefont {Jia}, \citenamefont {Kawamura}, \citenamefont {Ko},
  \citenamefont {Kokalj}, \citenamefont {K{\"u}{\c c}{\"u}kbenli},
  \citenamefont {Lazzeri}, \citenamefont {Marsili}, \citenamefont {Marzari},
  \citenamefont {Mauri}, \citenamefont {Nguyen}, \citenamefont {Nguyen},
  \citenamefont {{Otero-de-la-Roza}}, \citenamefont {Paulatto}, \citenamefont
  {Ponc{\'e}}, \citenamefont {Rocca}, \citenamefont {Sabatini}, \citenamefont
  {Santra}, \citenamefont {Schlipf}, \citenamefont {Seitsonen}, \citenamefont
  {Smogunov}, \citenamefont {Timrov}, \citenamefont {Thonhauser}, \citenamefont
  {Umari}, \citenamefont {Vast}, \citenamefont {Wu},\ and\ \citenamefont
  {Baroni}}]{giannozziAdvancedCapabilitiesMaterials2017}%
  \BibitemOpen
  \bibfield  {author} {\bibinfo {author} {\bibfnamefont {P.}~\bibnamefont
  {Giannozzi}}, \bibinfo {author} {\bibfnamefont {O.}~\bibnamefont
  {Andreussi}}, \bibinfo {author} {\bibfnamefont {T.}~\bibnamefont {Brumme}},
  \bibinfo {author} {\bibfnamefont {O.}~\bibnamefont {Bunau}}, \bibinfo
  {author} {\bibfnamefont {M.~B.}\ \bibnamefont {Nardelli}}, \bibinfo {author}
  {\bibfnamefont {M.}~\bibnamefont {Calandra}}, \bibinfo {author}
  {\bibfnamefont {R.}~\bibnamefont {Car}}, \bibinfo {author} {\bibfnamefont
  {C.}~\bibnamefont {Cavazzoni}}, \bibinfo {author} {\bibfnamefont
  {D.}~\bibnamefont {Ceresoli}}, \bibinfo {author} {\bibfnamefont
  {M.}~\bibnamefont {Cococcioni}}, \bibinfo {author} {\bibfnamefont
  {N.}~\bibnamefont {Colonna}}, \bibinfo {author} {\bibfnamefont
  {I.}~\bibnamefont {Carnimeo}}, \bibinfo {author} {\bibfnamefont {A.~D.}\
  \bibnamefont {Corso}}, \bibinfo {author} {\bibfnamefont {S.}~\bibnamefont
  {de~Gironcoli}}, \bibinfo {author} {\bibfnamefont {P.}~\bibnamefont
  {Delugas}}, \bibinfo {author} {\bibfnamefont {R.~A.}\ \bibnamefont
  {DiStasio}}, \bibinfo {author} {\bibfnamefont {A.}~\bibnamefont {Ferretti}},
  \bibinfo {author} {\bibfnamefont {A.}~\bibnamefont {Floris}}, \bibinfo
  {author} {\bibfnamefont {G.}~\bibnamefont {Fratesi}}, \bibinfo {author}
  {\bibfnamefont {G.}~\bibnamefont {Fugallo}}, \bibinfo {author} {\bibfnamefont
  {R.}~\bibnamefont {Gebauer}}, \bibinfo {author} {\bibfnamefont
  {U.}~\bibnamefont {Gerstmann}}, \bibinfo {author} {\bibfnamefont
  {F.}~\bibnamefont {Giustino}}, \bibinfo {author} {\bibfnamefont
  {T.}~\bibnamefont {Gorni}}, \bibinfo {author} {\bibfnamefont
  {J.}~\bibnamefont {Jia}}, \bibinfo {author} {\bibfnamefont {M.}~\bibnamefont
  {Kawamura}}, \bibinfo {author} {\bibfnamefont {H.-Y.}\ \bibnamefont {Ko}},
  \bibinfo {author} {\bibfnamefont {A.}~\bibnamefont {Kokalj}}, \bibinfo
  {author} {\bibfnamefont {E.}~\bibnamefont {K{\"u}{\c c}{\"u}kbenli}},
  \bibinfo {author} {\bibfnamefont {M.}~\bibnamefont {Lazzeri}}, \bibinfo
  {author} {\bibfnamefont {M.}~\bibnamefont {Marsili}}, \bibinfo {author}
  {\bibfnamefont {N.}~\bibnamefont {Marzari}}, \bibinfo {author} {\bibfnamefont
  {F.}~\bibnamefont {Mauri}}, \bibinfo {author} {\bibfnamefont {N.~L.}\
  \bibnamefont {Nguyen}}, \bibinfo {author} {\bibfnamefont {H.-V.}\
  \bibnamefont {Nguyen}}, \bibinfo {author} {\bibfnamefont {A.}~\bibnamefont
  {{Otero-de-la-Roza}}}, \bibinfo {author} {\bibfnamefont {L.}~\bibnamefont
  {Paulatto}}, \bibinfo {author} {\bibfnamefont {S.}~\bibnamefont {Ponc{\'e}}},
  \bibinfo {author} {\bibfnamefont {D.}~\bibnamefont {Rocca}}, \bibinfo
  {author} {\bibfnamefont {R.}~\bibnamefont {Sabatini}}, \bibinfo {author}
  {\bibfnamefont {B.}~\bibnamefont {Santra}}, \bibinfo {author} {\bibfnamefont
  {M.}~\bibnamefont {Schlipf}}, \bibinfo {author} {\bibfnamefont {A.~P.}\
  \bibnamefont {Seitsonen}}, \bibinfo {author} {\bibfnamefont {A.}~\bibnamefont
  {Smogunov}}, \bibinfo {author} {\bibfnamefont {I.}~\bibnamefont {Timrov}},
  \bibinfo {author} {\bibfnamefont {T.}~\bibnamefont {Thonhauser}}, \bibinfo
  {author} {\bibfnamefont {P.}~\bibnamefont {Umari}}, \bibinfo {author}
  {\bibfnamefont {N.}~\bibnamefont {Vast}}, \bibinfo {author} {\bibfnamefont
  {X.}~\bibnamefont {Wu}},\ and\ \bibinfo {author} {\bibfnamefont
  {S.}~\bibnamefont {Baroni}},\ }\bibfield  {title} {\bibinfo {title} {Advanced
  capabilities for materials modelling with {{Quantum ESPRESSO}}},\ }\href
  {https://doi.org/10.1088/1361-648X/aa8f79} {\bibfield  {journal} {\bibinfo
  {journal} {J. Phys.: Condens. Matter}\ }\textbf {\bibinfo {volume} {29}},\
  \bibinfo {pages} {465901} (\bibinfo {year} {2017})}\BibitemShut {NoStop}%
\bibitem [{\citenamefont {Baroni}\ \emph {et~al.}(2001)\citenamefont {Baroni},
  \citenamefont {{de Gironcoli}}, \citenamefont {Dal~Corso},\ and\
  \citenamefont {Giannozzi}}]{baroniPhononsRelatedCrystal2001}%
  \BibitemOpen
  \bibfield  {author} {\bibinfo {author} {\bibfnamefont {S.}~\bibnamefont
  {Baroni}}, \bibinfo {author} {\bibfnamefont {S.}~\bibnamefont {{de
  Gironcoli}}}, \bibinfo {author} {\bibfnamefont {A.}~\bibnamefont
  {Dal~Corso}},\ and\ \bibinfo {author} {\bibfnamefont {P.}~\bibnamefont
  {Giannozzi}},\ }\bibfield  {title} {\bibinfo {title} {Phonons and related
  crystal properties from density-functional perturbation theory},\ }\href
  {https://doi.org/10.1103/RevModPhys.73.515} {\bibfield  {journal} {\bibinfo
  {journal} {Rev. Mod. Phys.}\ }\textbf {\bibinfo {volume} {73}},\ \bibinfo
  {pages} {515} (\bibinfo {year} {2001})}\BibitemShut {NoStop}%
\bibitem [{\citenamefont {Clay}\ \emph {et~al.}(2016)\citenamefont {Clay},
  \citenamefont {Holzmann}, \citenamefont {Ceperley},\ and\ \citenamefont
  {Morales}}]{PhysRevB.93.035121}%
  \BibitemOpen
  \bibfield  {author} {\bibinfo {author} {\bibfnamefont {R.~C.}\ \bibnamefont
  {Clay}}, \bibinfo {author} {\bibfnamefont {M.}~\bibnamefont {Holzmann}},
  \bibinfo {author} {\bibfnamefont {D.~M.}\ \bibnamefont {Ceperley}},\ and\
  \bibinfo {author} {\bibfnamefont {M.~A.}\ \bibnamefont {Morales}},\
  }\bibfield  {title} {\bibinfo {title} {Benchmarking density functionals for
  hydrogen-helium mixtures with quantum monte carlo: Energetics, pressures, and
  forces},\ }\href {https://doi.org/10.1103/PhysRevB.93.035121} {\bibfield
  {journal} {\bibinfo  {journal} {Phys. Rev. B}\ }\textbf {\bibinfo {volume}
  {93}},\ \bibinfo {pages} {035121} (\bibinfo {year} {2016})}\BibitemShut
  {NoStop}%
\bibitem [{\citenamefont {Dynes}(1972)}]{dynesMcMillanEquationTc1972}%
  \BibitemOpen
  \bibfield  {author} {\bibinfo {author} {\bibfnamefont {R.~C.}\ \bibnamefont
  {Dynes}},\ }\bibfield  {title} {\bibinfo {title} {{{McMillan}}'s equation and
  the {{Tc}} of superconductors},\ }\href
  {https://doi.org/10.1016/0038-1098(72)90603-5} {\bibfield  {journal}
  {\bibinfo  {journal} {Solid State Communications}\ }\textbf {\bibinfo
  {volume} {10}},\ \bibinfo {pages} {615} (\bibinfo {year} {1972})}\BibitemShut
  {NoStop}%
\bibitem [{\citenamefont {Allen}\ and\ \citenamefont
  {Dynes}(1975)}]{allenTransitionTemperatureStrongcoupled1975}%
  \BibitemOpen
  \bibfield  {author} {\bibinfo {author} {\bibfnamefont {P.~B.}\ \bibnamefont
  {Allen}}\ and\ \bibinfo {author} {\bibfnamefont {R.~C.}\ \bibnamefont
  {Dynes}},\ }\bibfield  {title} {\bibinfo {title} {Transition temperature of
  strong-coupled superconductors reanalyzed},\ }\href
  {https://doi.org/10.1103/PhysRevB.12.905} {\bibfield  {journal} {\bibinfo
  {journal} {Phys. Rev. B}\ }\textbf {\bibinfo {volume} {12}},\ \bibinfo
  {pages} {905} (\bibinfo {year} {1975})}\BibitemShut {NoStop}%
\end{thebibliography}%

\newpage

\appendix
%\section*{Supplementary Information}
\setcounter{figure}{0}
\counterwithin{figure}{section}
\renewcommand{\thefigure}{S\arabic{figure}}
\setcounter{table}{0}
\counterwithin{table}{section}
\renewcommand{\thetable}{S\arabic{table}}

\newpage

\begin{figure*}[htp]
\centering
\includegraphics[width=0.9\textwidth]{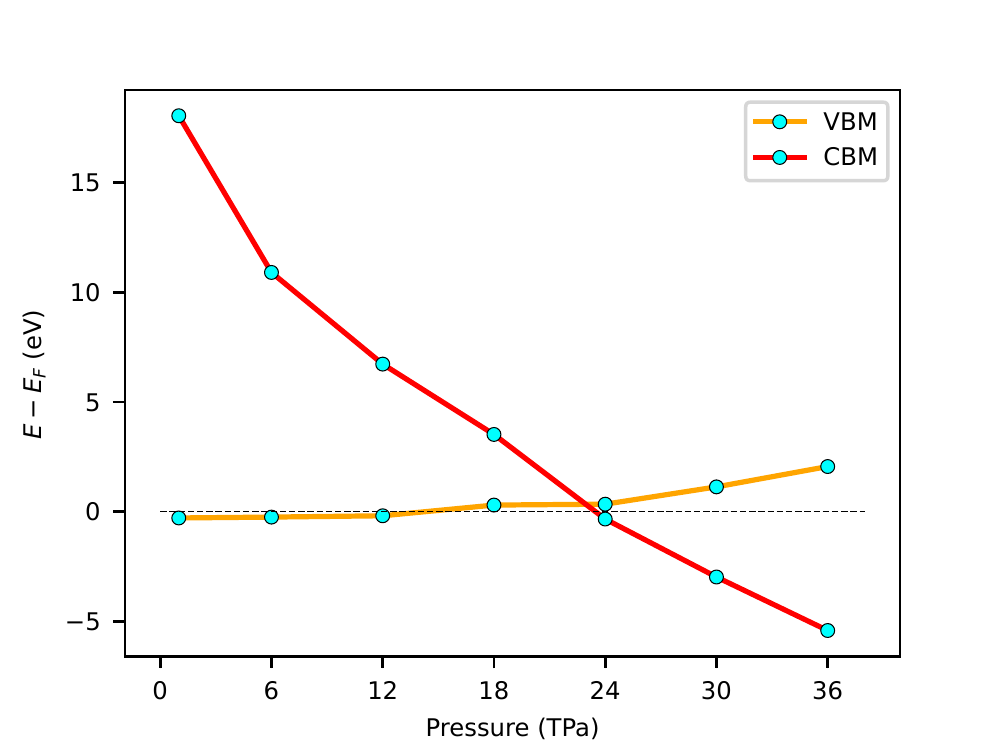}
\caption{Evolution of VBM and CBM levels in hcp $^{4}$He expressed as a function of compression and calculated with the hybrid B3LYP functional. The dashed line represents the Fermi energy level. The crossing of the two curves indicates metallization. The CBM level (red curve), which presents full $p$-like electronic character, is efficiently tuned by pressure. The VBM level (orange curve), which presents mixed $s$--$p$ like electronic character, hardly reacts to pressure below the predicted metallization point of $24$~TPa.}
\label{fig:orbit_compression}
\end{figure*}

\newpage

\begin{figure*}[htp]
\centering
\includegraphics[width=0.9\textwidth]{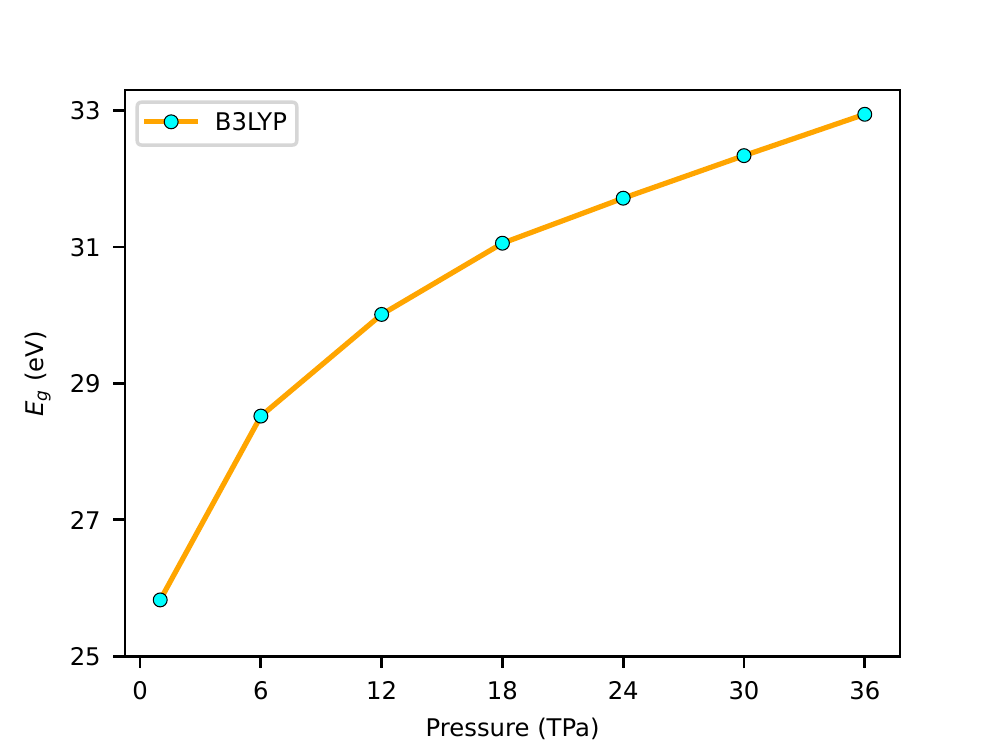}
\caption{Direct band gap at the reciprocal space point $\Lambda$ point calculated in hcp $^{4}$He as a function of pressure. The results were obtained with the hybrid B3LYP functional.}
\label{fig:direct_bandgap}
\end{figure*}

\newpage

\begin{figure*}[htp]
\centering
\includegraphics[width=1.0\textwidth]{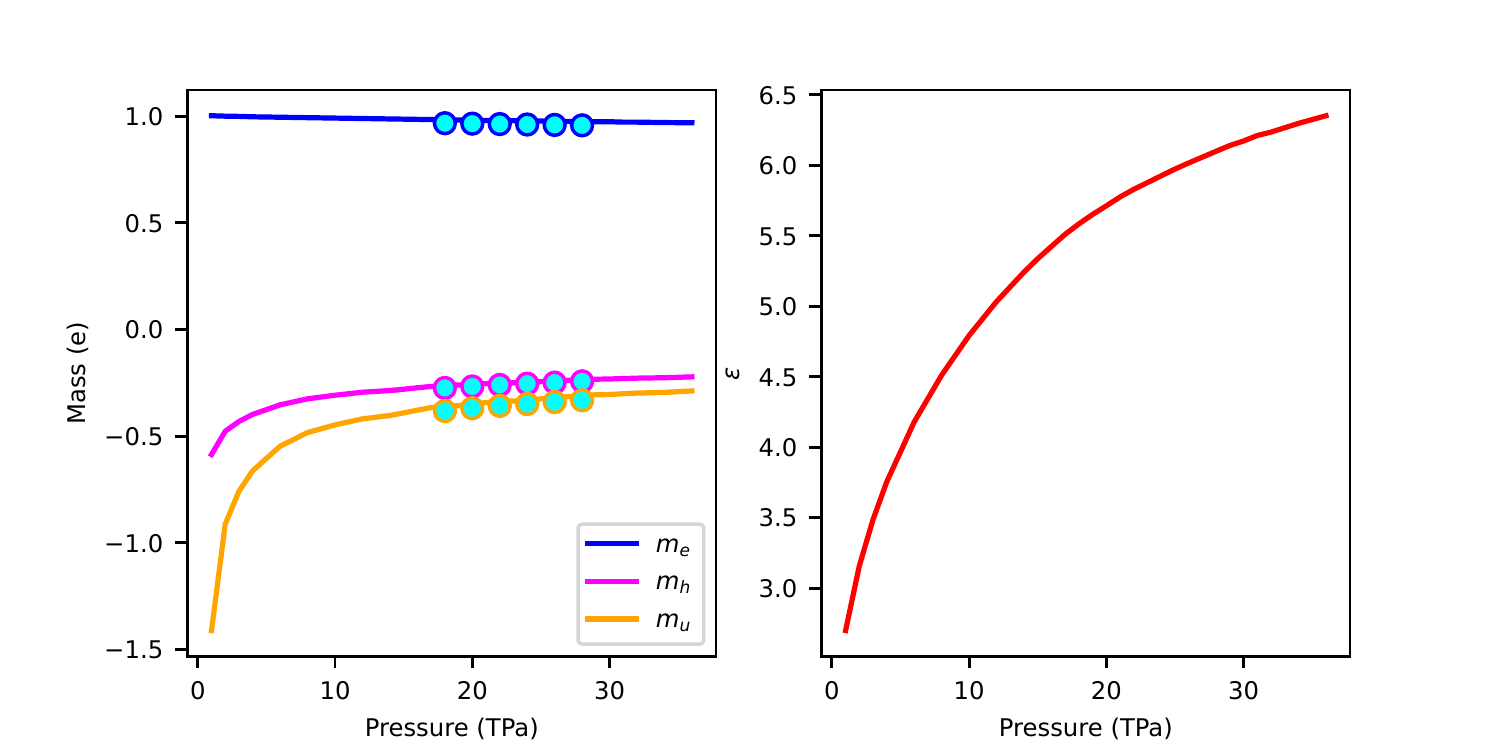}
\caption{Pressure evolution of the effective mass of electrons and holes and the relative dielectric constant in hcp $^{4}$He. Solid curves are calculated with the semi-local PBEsol functional while the cyan points correspond to hybrid B3LYP calculations. It is found that the choice of the DFT functional does not appreciably affect the estimated effective masses of the electrons and holes as obtained from the curvature of the valence and conduction bands respectively. Thus, the choice of the DFT functional only influences noticeably the band gap.}
\label{fig:mass}
\end{figure*}

\newpage

\begin{figure*}[htp]
\centering
\includegraphics[width=1.0\textwidth]{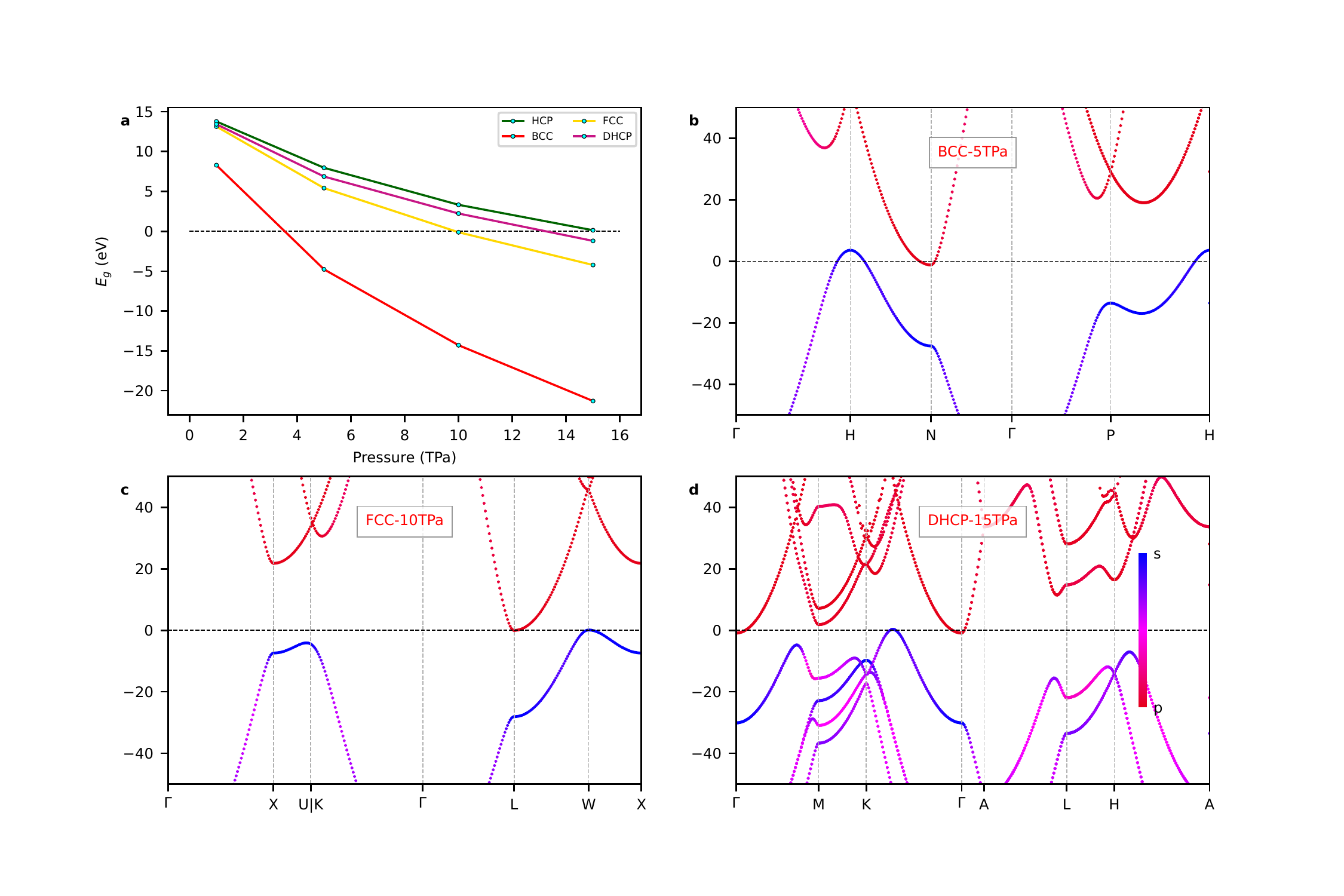}
\caption{Electronic band-structure properties calculated for $^{4}$He considering other possible crystal structures, namely, body-centered cubic (bcc), face-centered cubic (fcc), and double hexagonal closed packed (dhcp). \textbf{a}~Band gap expressed as a function of pressure and calculated with the semi-local PBEsol functional. \textbf{b-d}~Projected electronic band structure of metallic helium calculated in the bcc, fcc and dhcp phases, respectively.}
\label{fig:bcc}
\end{figure*}

\newpage

\begin{figure*}[htp]
\centering
\includegraphics[width=1.0\textwidth]{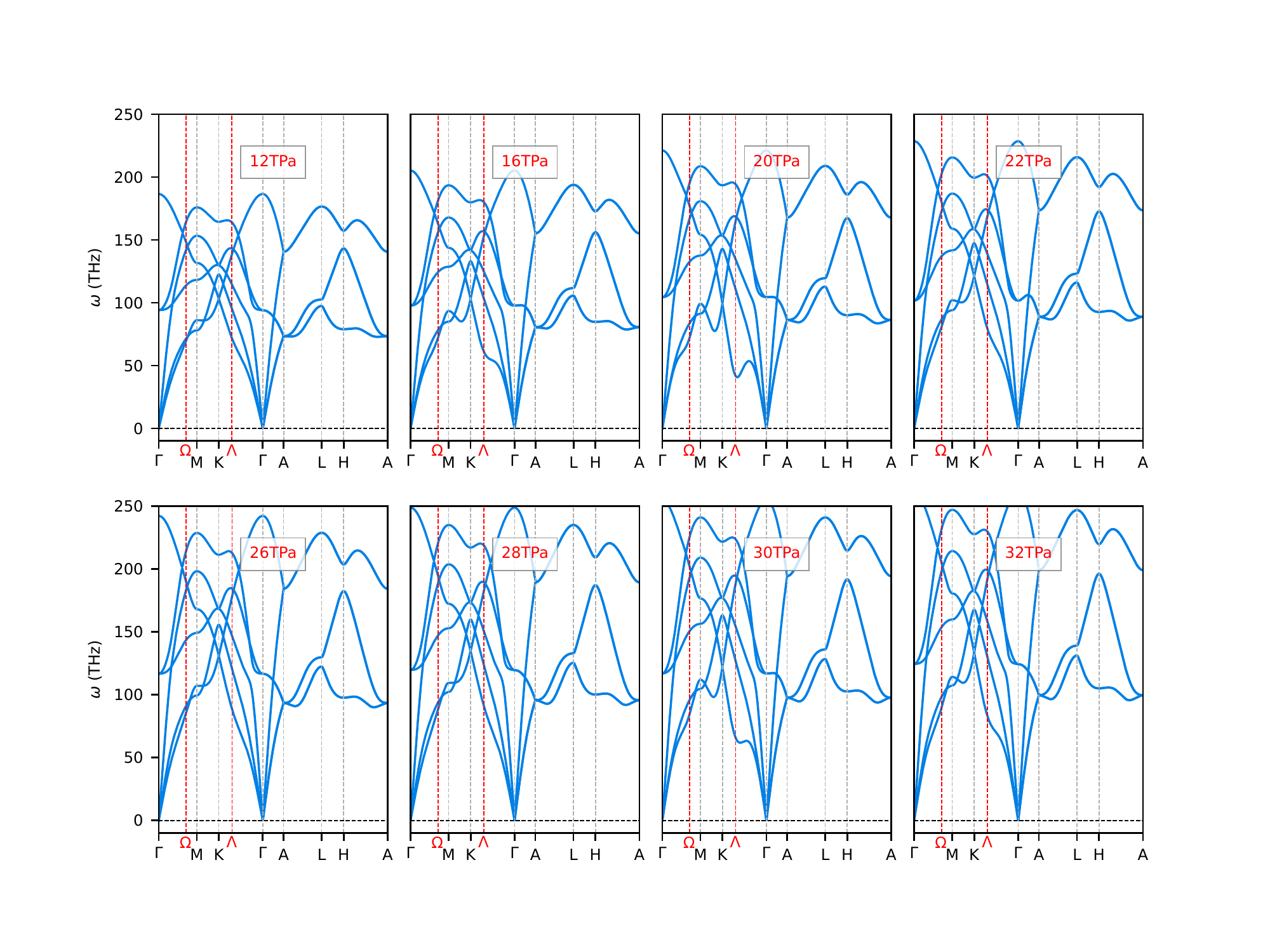}
\caption{The phonon spectrum of hcp $^{4}$He calculated at different pressures using the semi-local PBEsol functional in VASP. The $\Lambda$ and $\Omega$ reciprocal space points, which are not a highly symmetric, correspond to the VBM and the secondary VBM levels, respectively.}
\label{fig:phonon}
\end{figure*}

\newpage

\begin{figure*}[htp]
\centering
\includegraphics[width=1.0\textwidth]{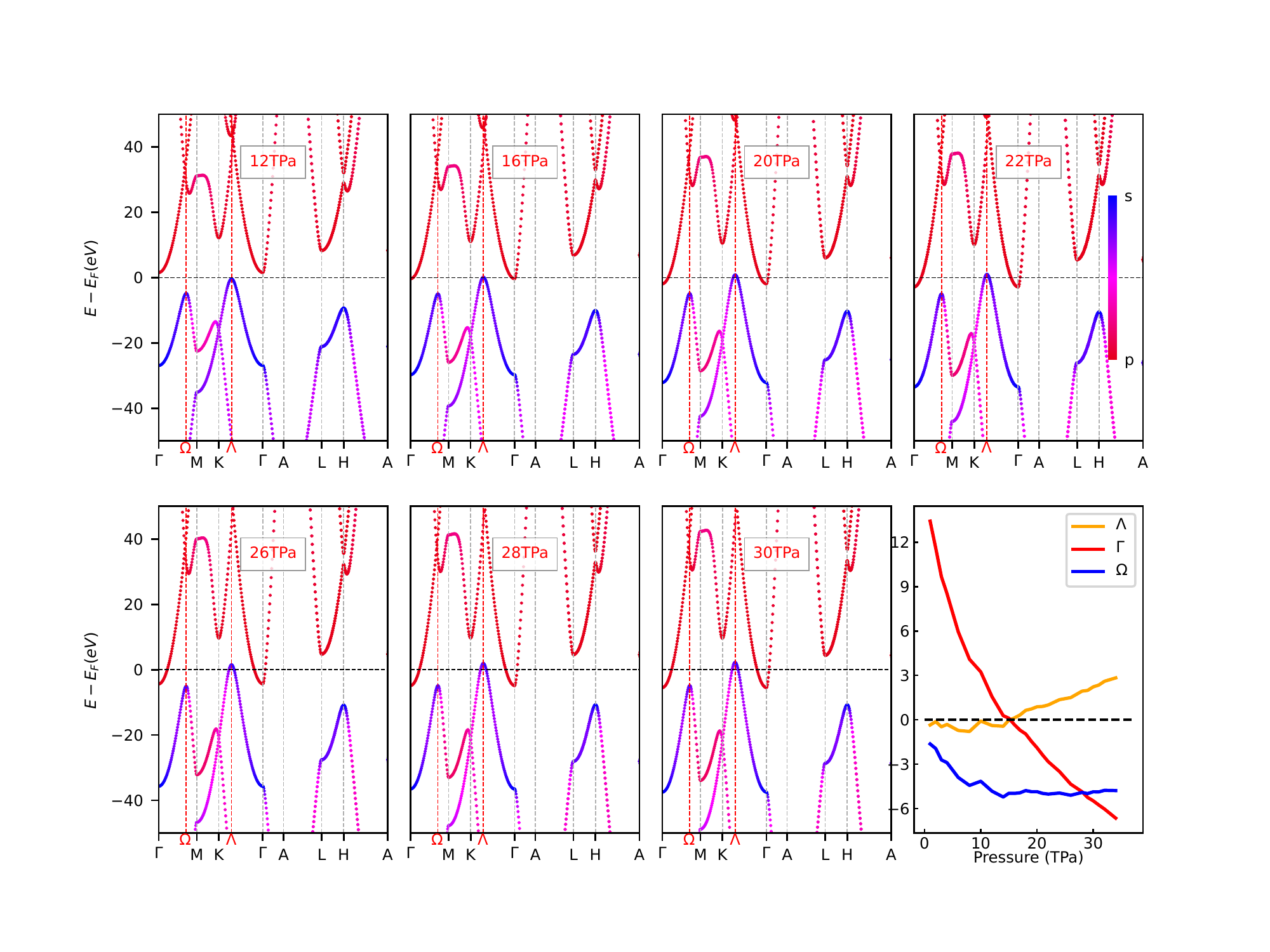}
\caption{Projected electronic band structure of hcp $^{4}$He at different pressures calculated with the semi-local PBEsol functional. The $\Lambda$ and $\Omega$ reciprocal space points, which are not highly symmetric, correspond to the VBM and secondary VBM levels, respectively. A red-magenta-blue color scale is used to represent the dominant character of the valence and conduction bands; red denotes $s$-like character and blue $p$-like. The last panel shows the pressure-induced energy evolution of the relevant reciprocal space points $\Lambda$, $\Gamma$ and $\Omega$ point, as referred to the Fermi surface.}
\label{fig:PBEsol_band}
\end{figure*}

\newpage

\begin{figure*}[htp]
\includegraphics[angle=0,width=0.9\textwidth]{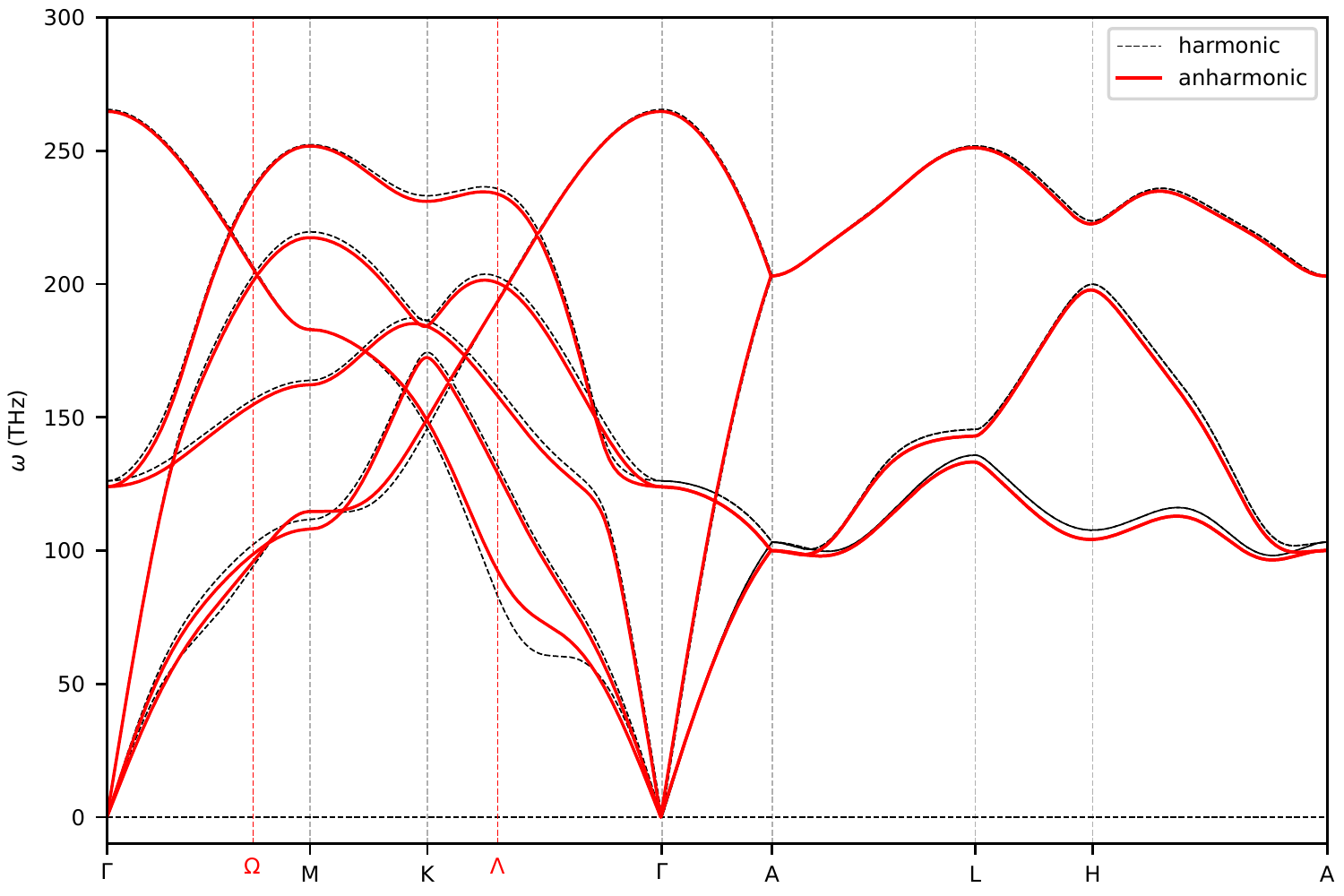}
\caption{\label{fig:sscha-2} Quantum anharmonic effects on the phonon spectrum of ultra-compressed hcp $^4$He. The comparison between the harmonic (dashed black lines) and the anharmonic (solid red lines) phonon spectra. The harmonic phonon spectra are taken from the DFPT calculations and the anharmonic phonon spectra are calculated from the free energy Hessian dynamical matrix $\mathbcal{D}^F$ up to the third-order force constants.}
\end{figure*}

\newpage

\begin{figure*}[htp]
\centering
\includegraphics[width=1.0\textwidth]{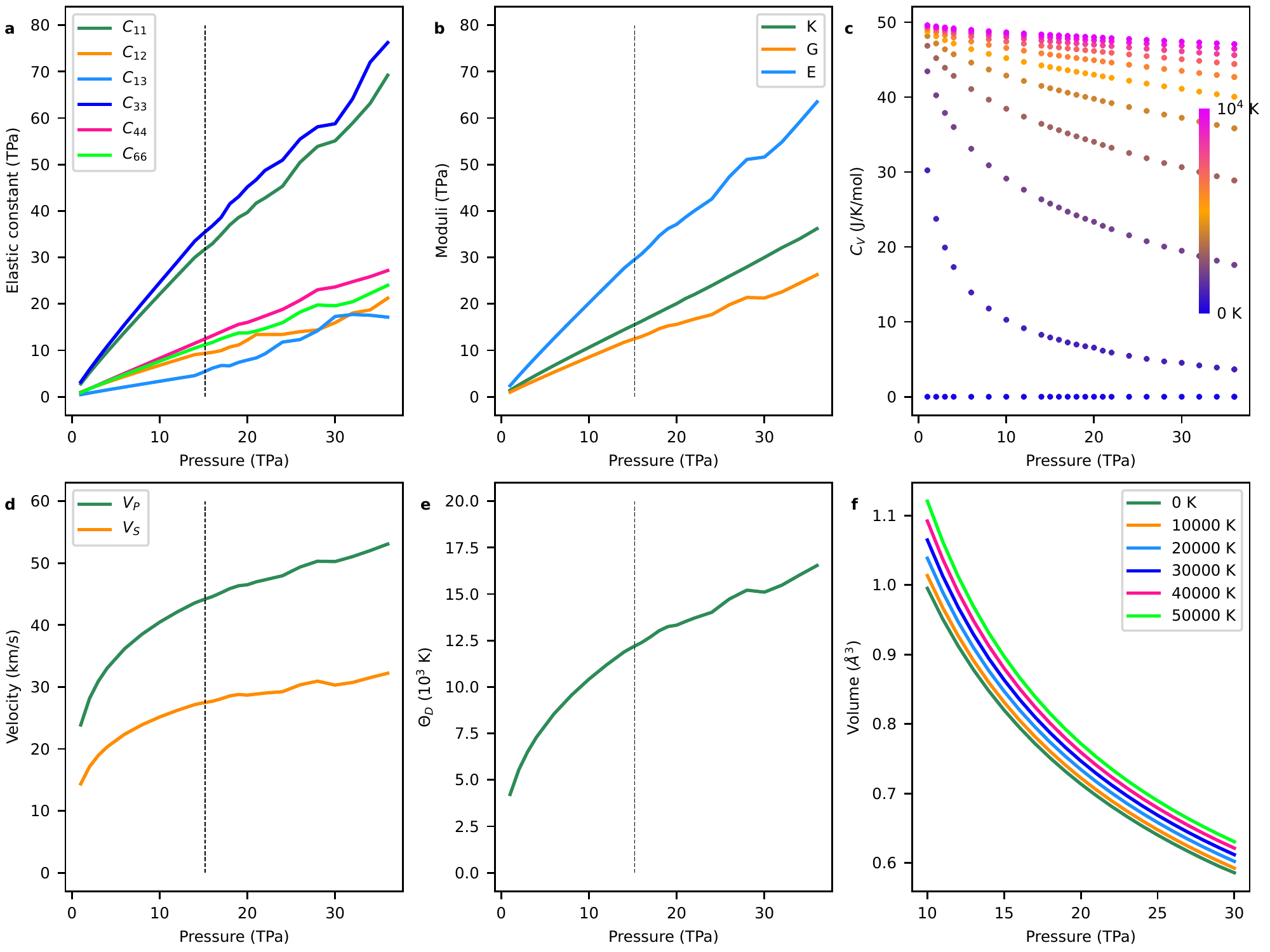}
\caption{Elastic properties of hcp $^{4}$He expressed as a function of pressure and calculated with the semi-local PBEsol functional. \textbf{a}~Elastic constants $\lbrace C_{ij} \rbrace$. \textbf{b}~Bulk modulus ($K$), shear modulus ($G$), and Young's modulus ($E$). \textbf{c}~The heat capacity ($C_V$) of the crystal calculated at fixed volume $V$. $C_{V}$ was calculated within the temperature interval $0 \le T \le 10,000$~K. \textbf{d}~Elastic sound wave velocity: longitudinal wave velocity ($v_p$) and transverse wave velocity ($v_s$). \textbf{e}~Debye temperature ($\Theta_D$) estimated for metallic helium. \textbf{f}~Temperature-dependence of the equation of state of hcp $^4$He as estimated with the quasi-harmonic approximation. 
The vertical dashed line in the figures marks the metallization pressure (i.e., $15$~TPa as obtained with the semi-local PBEsol functional). In the insulating phase, the pressure-dependence of the elastic constants and moduli are almost linear. In the metallic state, lattice distortions occur and those curves depart from the pressure-dependent linear behaviour; there is also a crossing between the $C_{12}$ and $C_{13}$ lines.}
\label{fig:cij}
\end{figure*}

\newpage

\begin{figure*}[htp]
\centering
\includegraphics[width=1.0\textwidth]{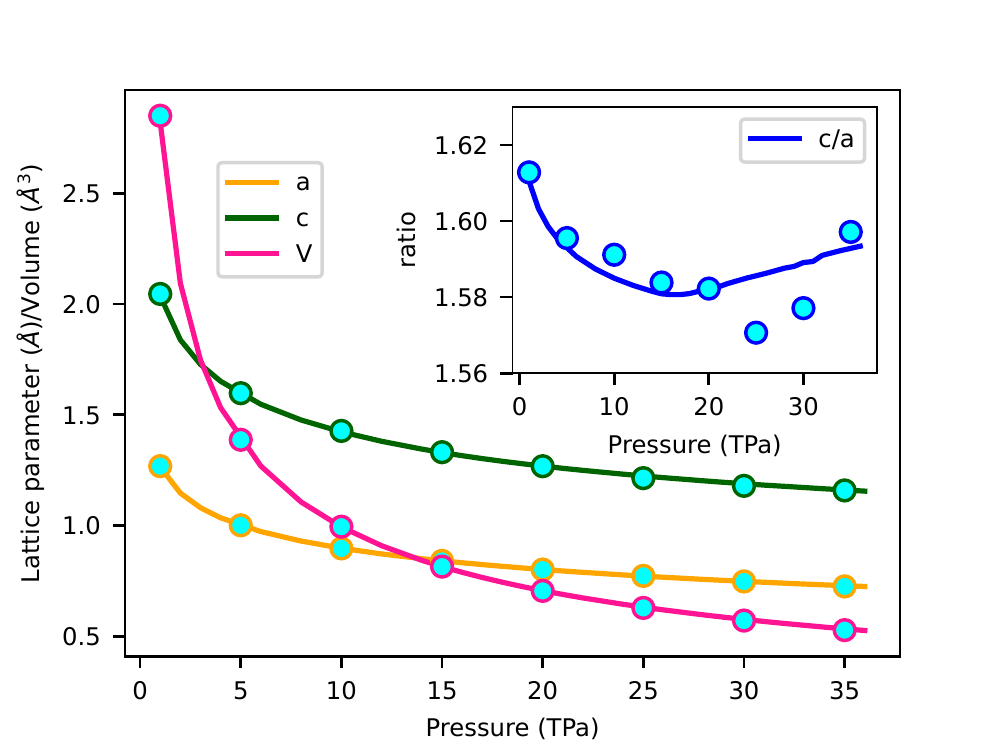}
\caption{Comparison of the equation of state of hcp $^{4}$He obtained with two different DFT software: Quantum Expresso (QE, circles) and VASP (curves). The length of the $a$ and $b$ lattice parameters and total volume expressed as a function of pressure and calculated with the QE and VASP codes agree remarkably well. The agreement between the two DFT codes on the value of the $c/a$ ratio, however, is not satisfactory within the pressure range $20 \le P \le 30$~TPa.}
\label{fig:eos}
\end{figure*}

\newpage

\begin{figure*}[htp]
\centering
\includegraphics[width=1.0\textwidth]{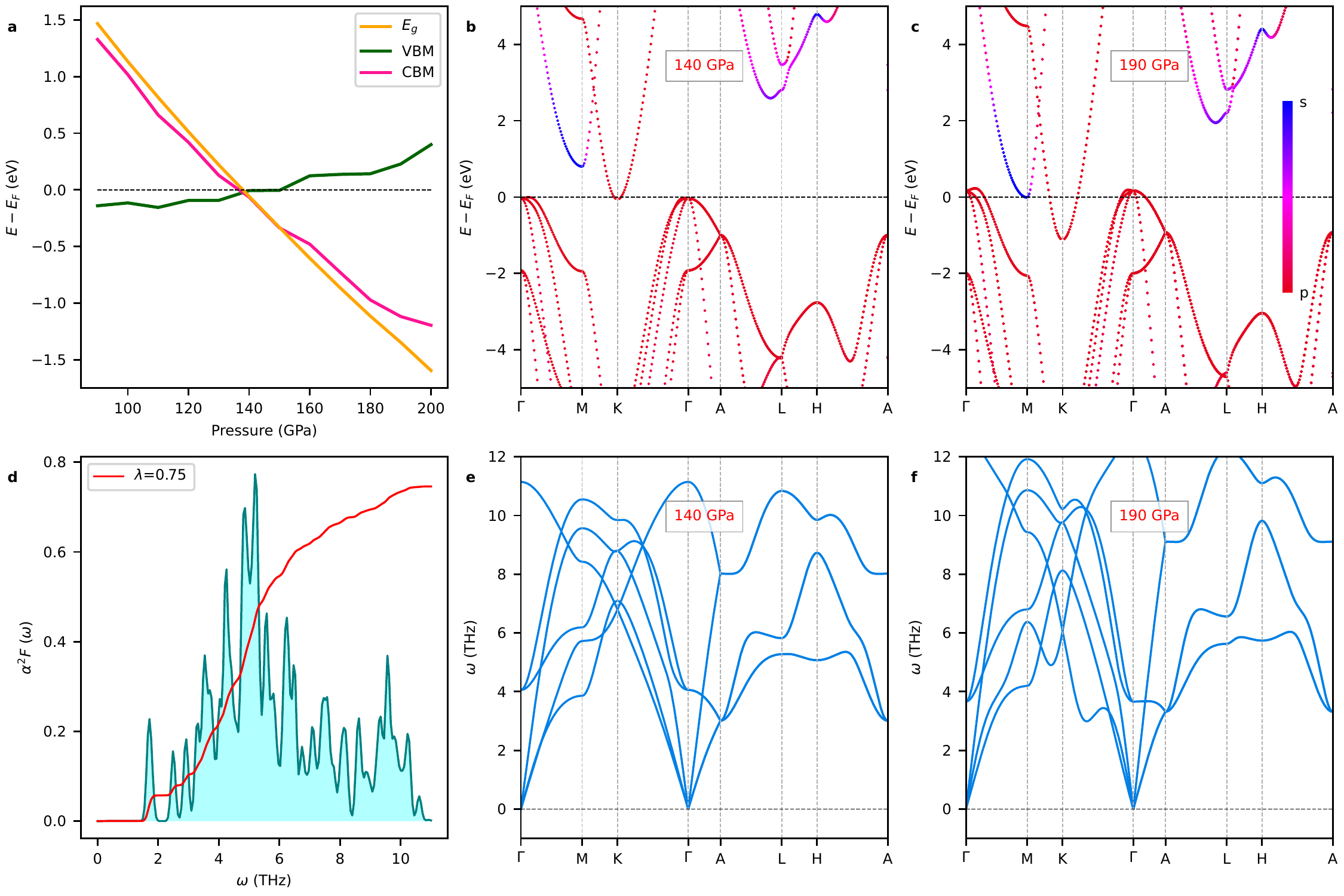}
\caption{Excitonic insulator state and superconductivity in solid hcp xenon. \textbf{a}~Band gap, VBM and CBM pressure-induced evolution. \textbf{b-c}~Projected electronic band structure and \textbf{e-f}~Phonon spectrum of hcp Xe at $140$ and $190$~GPa calculated with the semi-local PBEsol functional. \textbf{d}~Eliashberg spectral function, $\alpha^2F(\omega)$, estimated at $140$~GPa along with the integrated electron–phonon coupling constant $\lambda$. In analogy to helium, solid xenon also maintains the hcp crystal structure at high pressures (i.e., $\sim 100$~GPa). For the electronic band structure and phonon spectrum calculations, we considered the electrons $5s^25p^6$ as valence states and used the semi-local PBEsol DFT functional with a plane wave cutoff of $300$~eV. For EPC calculations, we used ultrasoft pseudopotentials, a plane-wave cut-off energy of $50$~Ry for the kinetic energy and of $500$~Ry for the charge density. We adopted a dense and shifted $k$-point mesh of $16 \times 16 \times 8$ for the self-consistent calculation and then a denser $32 \times 32 \times 16$ grid for further EPC calculations. As it is shown the figure, solid xenon becomes metallic at an experimentally accessible pressure of $140$~GPa. Its band gap is also indirect with the VBM located at the $\Gamma$ point and the CBM at $K$ point. VBM and CBM exhibit pure electronic $p$-like character while the secondary CBM $s$-like. No phonon softening occurs at $140$~GPa as the band gap closure involves $p$-$p$ orbital interactions. Meanwhile, phonon softening is observed at $190$~GPa upon closure of the secondary band gap which involves $s$-$p$ orbital interactions, in analogy to solid helium. The Eliashberg spectral function calculated at $140$~GPa shows that the electron phonon coupling strength ($\lambda = 0.75$) mainly stems from the low-frequency region (i.e., $2$--$6$~THz). In this case, the resulting critical superconducting temperature is $10$~K.}
\label{fig:Xe}
\end{figure*}

\newpage

\begin{figure*}[htp]
\centering
\includegraphics[width=1.0\textwidth]{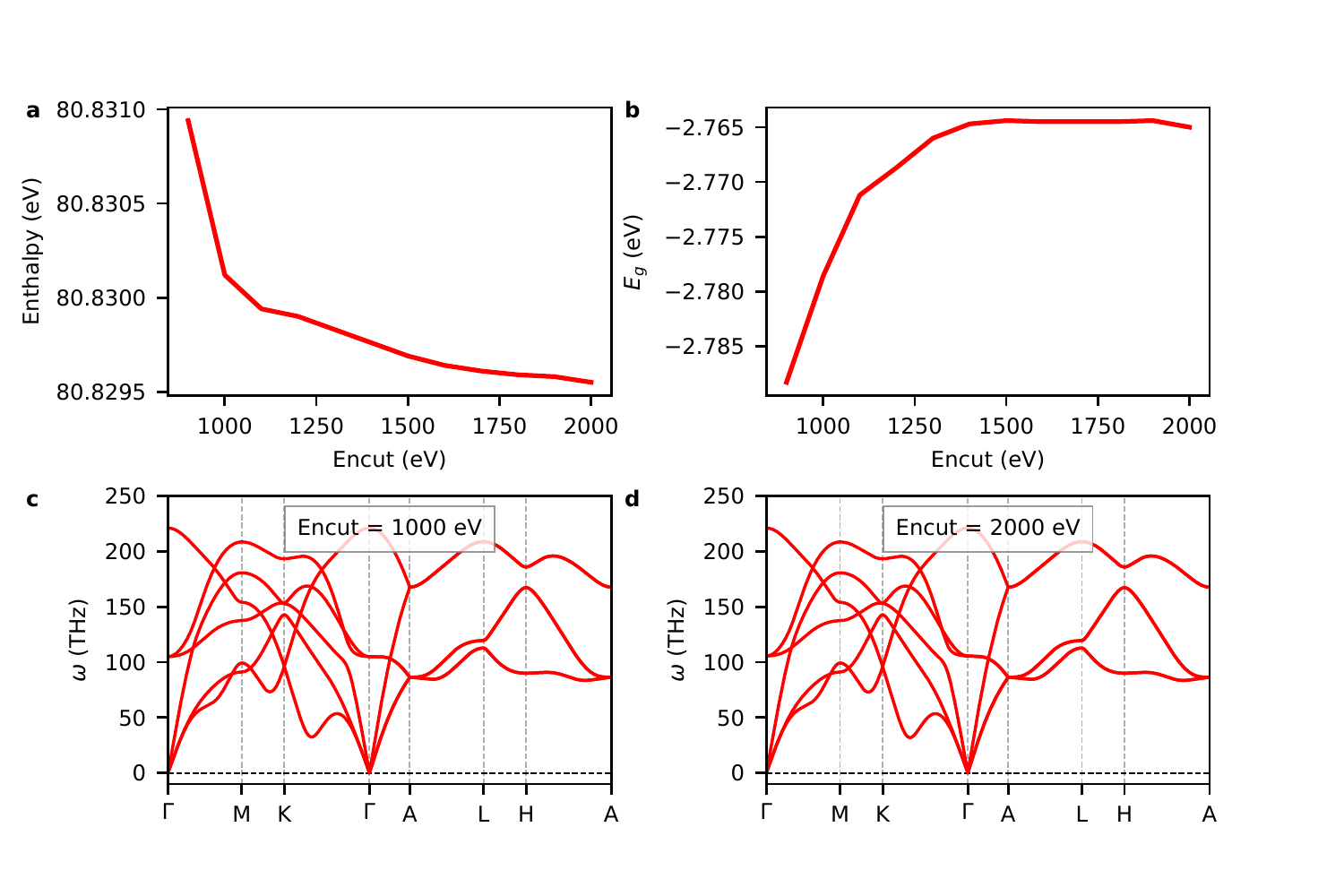}
\caption{Energy cutoff tests performed with the VASP code and the semi-local PBEsol functional. Results are shown for the \textbf{a}~enthalpy, \textbf{b}~band gap and \textbf{c-d}~phonon spectrum of solid helium at $20$~TPa. It is found that an energy cutoff of $1500$~eV guarantees convergence in all the quantities to the desired level of accuracy.}
\label{fig:encut_test}
\end{figure*}

\newpage

\begin{figure*}[htp]
\centering
\includegraphics[width=0.9\textwidth]{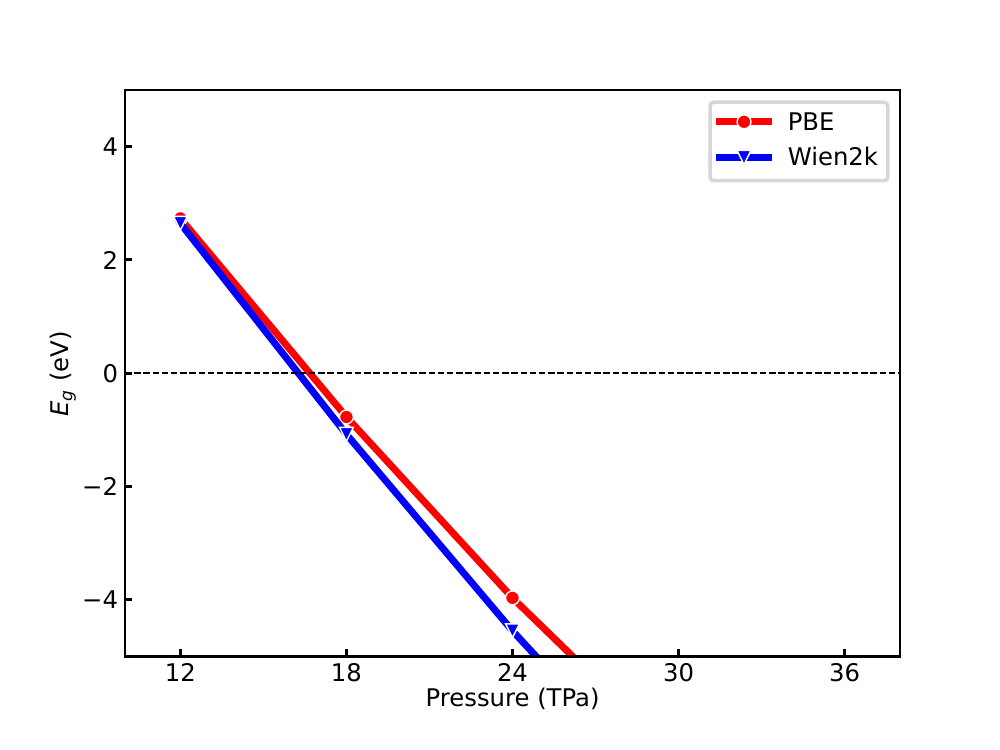}
\caption{Comparison of the PBE band gap evolution obtained with a PAW pseudopotential (red line, results obtained with the VASP code) and an all-electron approach (blue line, results obtained with the WIEN2K code). The full-potential (linearized) augmented plane-wave method is consistent with the employed pseudopotential PBE approach when evaluating band structure closure under such a high pressure.}
\label{fig:wein2k}
\end{figure*}

\newpage

\begin{figure*}[htp]
\centering
\includegraphics[width=0.9\textwidth]{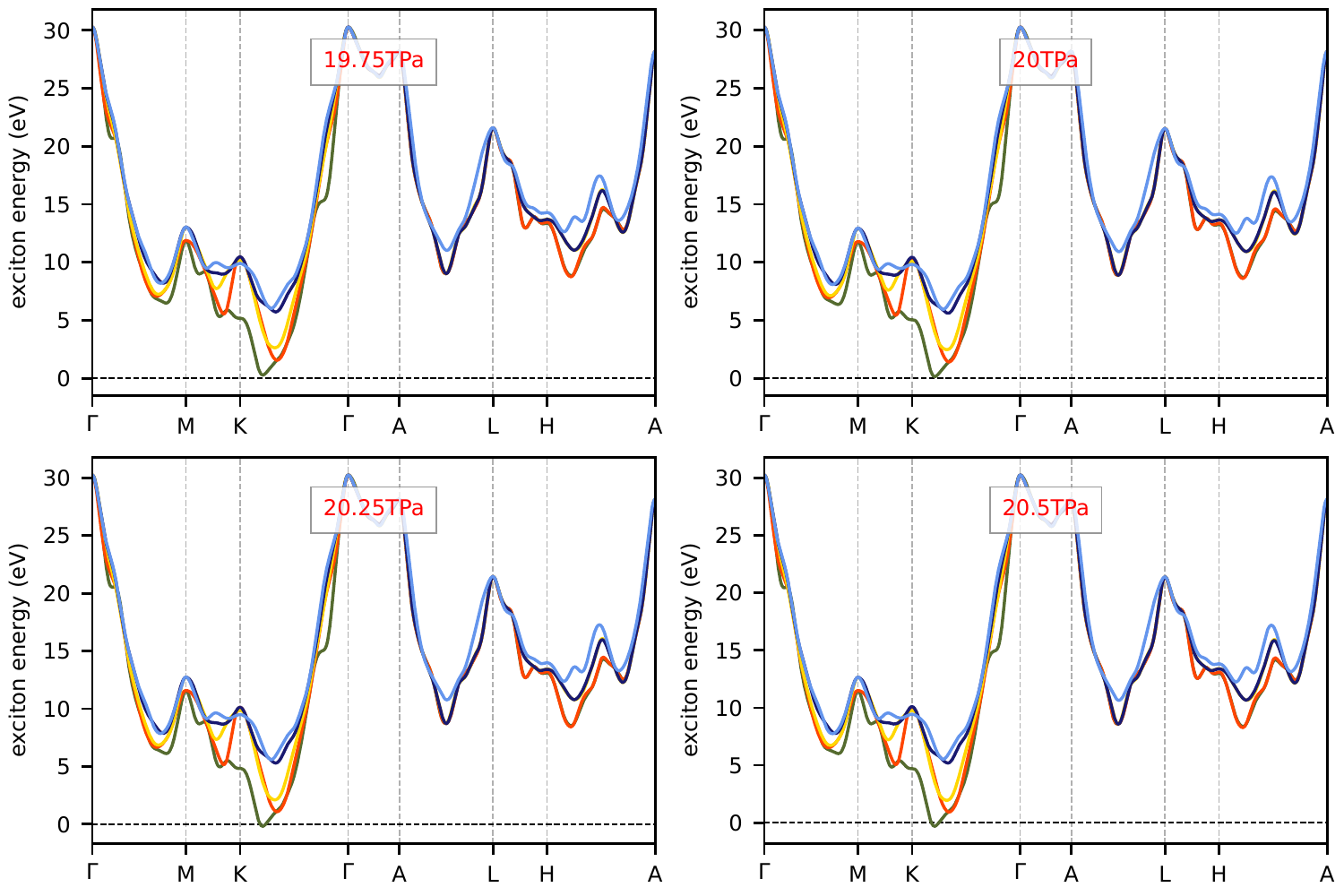}
\caption{Pressure evolution of the excitonic energy obtained with GW calculations. The interpolated exciton energies are mapped into different ${\bf q}$-point with colorful curves indicating different bands. The exciton in path $K$--$\Gamma$ has the lowest energy with a value of $-0.31$~eV at $20.25$~TPa, which is negative, indicating that the exciton state is bound.}
\label{fig:exciton}
\end{figure*}

\newpage

\begin{table*}[htp]
\centering
\caption{The EPC strength parameter $\lambda$, electronic density of states at the Fermi level $N(E_F)$ (per formula unit), logarithmic average phonon frequency $\omega_{log}$, and critical superconducting temperature for metallic hcp $^4$He at different pressures, $P$. The critical superconducting temperature was calculated with different approaches using $\mu^* = 0.10$ and $0.13$ (the latter case within parentheses). $T_{c}^{McM}$ was calculated with the McMillan formula (\ref{eq:McM}), $T_{c}^{AD}$ with the Allen-Dynes formula (\ref{eq:AD}), and $T_{c}^{mAD}$ with the modified Allen-Dynes formula (\ref{eq:mAD}).}
\label{table:tc}
\vspace{0.5cm}
\centering
\begin{tabular}{c c c c c c c}  
\hline
\hline
\quad $P$~(TPa) \quad & \quad $\lambda$  \quad & \quad $\omega_{log}$~(K) \quad & \quad $N(E_F)$ \quad & \quad $T_{c}^{McM}$~(K) \quad & \quad $T_{c}^{AD}$~(K) \quad & \quad  $T_{c}^{mAD}$~(K) \quad \quad \\
\hline
 20 & 0.45 & 3888 & 0.011294 & 30 (16) & 31 (16) & 32 (17) \\
 30 & 0.22 & 4614 & 0.047406 & 2 (1) & 2 (1) & 2 (1) \\
 40 & 0.38 & 4597 & 0.059410 & 16 (7) & 16 (7) & 17 (7) \\
 50 & 0.52 & 5182 & 0.057635 & 71 (44) & 73 (44) & 76 (47) \\ 
 100 & 0.52 & 6842 & 0.051989 & 97 (61) & 99 (62) & 103 (65) \\
\hline
\hline
\end{tabular}
\end{table*}

\newpage

\begin{table*}[htp]
\centering
\caption{The band gap ($E_g$), excitonic energy ($E_e$) and excitonic binding energy ($E_{bind}$) obtained with GW calculations at several different pressure points ($P$). Band gaps are directly computed from the valence band maximum (VBM) and conduction band minimum (CBM).}
\label{table:exciton}
\vspace{0.5cm}
\begin{tabular}{c c c c c c}  
\hline
\hline
\quad $P$~(TPa) \quad & \quad VBM~(eV)  \quad & \quad CBM~(eV) \quad & \quad $E_g$~(eV) \quad & \quad $E_e$~(eV) \quad & \quad  $E_{bind}$~(eV) \quad  \\
\hline
19.50&	-0.77697&	0.16674&	0.94371&	-- &	-- \\
19.75&	-0.16201&	0.61332&	0.77533&	0.27661&	0.49872 \\
20.00&	-0.08247&	0.53441&	0.61688&	0.13751&	0.47937 \\
20.25&	0.05131&	0.37315&	0.32184&	-0.18292&	0.50476 \\
20.50&	0.11935&	0.29978&	0.18043&	-0.31057&	0.49043 \\
20.75&	0.19766&	0.13734&	-0.06032&	-- &	-- \\
21.00&	0.28342&	0.01436&	-0.26906&	-- &	-- \\
\hline
\hline
\end{tabular}
\end{table*}

\newpage

\begin{table*}[htp]
\caption{Cutoff energy convergence tests. Electronic density of states at the Fermi level, $N(E_{F})$ (per formula unit), Fermi energy level, $E_{F}$, and EPC parameters calculated for the phonon at the wave vector $\Gamma$ with branch index $4$ and $q = (0.25,0.53,0)$ with branch index $1$ considering the Monkhorst–Pack (MP) grid scheme and a Gaussian broadening of $0.002$~Ry at $20$~TPa.}
\label{table:encut}
\vspace{0.25cm}
\centering
\begin{tabular}{c c c c c c}
\hline
\hline
$q$ & \quad Cutoff energy \quad & \quad $N(E_F)$ \quad &  $E_F$ \quad & \quad $\lambda_{q\nu}$ \quad & $\gamma_{q\nu}$ \quad \quad \\
    & (Ry) & (states/spin/Ry/formula) & (eV) &   & (GHz) \quad \quad \\
\hline
\multirow{4}{*}{(0,0,0)} & 100 & 0.011124 & 78.268714 & 39.9965 & 5191.84\\
& 200 & 0.011294 & 78.267634 & 40.5615 & 5344.54\\
& 300 & 0.011296 & 78.267626 & 40.5664 & 5345.93\\
& 400 & 0.011295 & 78.267614 & 40.5634 & 5345.21\\
\hline
\multirow{4}{*}{(0.25,0.43,0)} & 100 & 0.011124 & 78.268714 & 135.1200 & 8891.19\\
& 200 & 0.011294 & 78.267634 & 140.5567 & 9375.37\\
& 300 & 0.011296 & 78.267626 & 140.5894 & 9378.88\\
& 400 & 0.011295 & 78.267614 & 140.5688 & 9377.09\\
\hline
\hline
\end{tabular}
\end{table*}

\newpage

\begin{table*}[htp]
\caption{Test on the ${\bf k}$-mesh convergence for $\Gamma$ and $\Lambda$ points with a Gaussian broadening of $0.002$~Ry at a pressure of $20$~TPa.}
\label{table:kmesh}
\vspace{0.25cm}
\centering
\begin{tabular}{c c c}
\hline
\hline
${\bf k}$-mesh	 &   $\lambda_{\Gamma}$ & $\lambda_{\Lambda}$ \\
\hline
96$\times$96$\times$48   &	35.9302 &	0.0011 \\
112$\times$112$\times$56 &	35.1325 &	0.5302 \\
128$\times$128$\times$64 &	38.2727 &	66.0029 \\
144$\times$144$\times$72 &	33.4231 &	57.1476 \\
160$\times$160$\times$80 &	40.1960 &	113.5229 \\
176$\times$176$\times$88 &	39.1550 &	220.7670 \\
192$\times$192$\times$96 &	40.5855 &	139.7354 \\
% 208$\times$208$\times$104 &	42.3404 &	167.9472 \\
% 224$\times$224$\times$112 &	43.4753 &	152.0405 \\
% 240$\times$240$\times$120 &	44.1517 &	132.0684 \\
256$\times$256$\times$128 &	44.5449 &	141.41328 \\
\hline
\hline
\end{tabular}
\end{table*}

\newpage

\begin{figure*}[htp]
\centering
\includegraphics[width=0.8\textwidth]{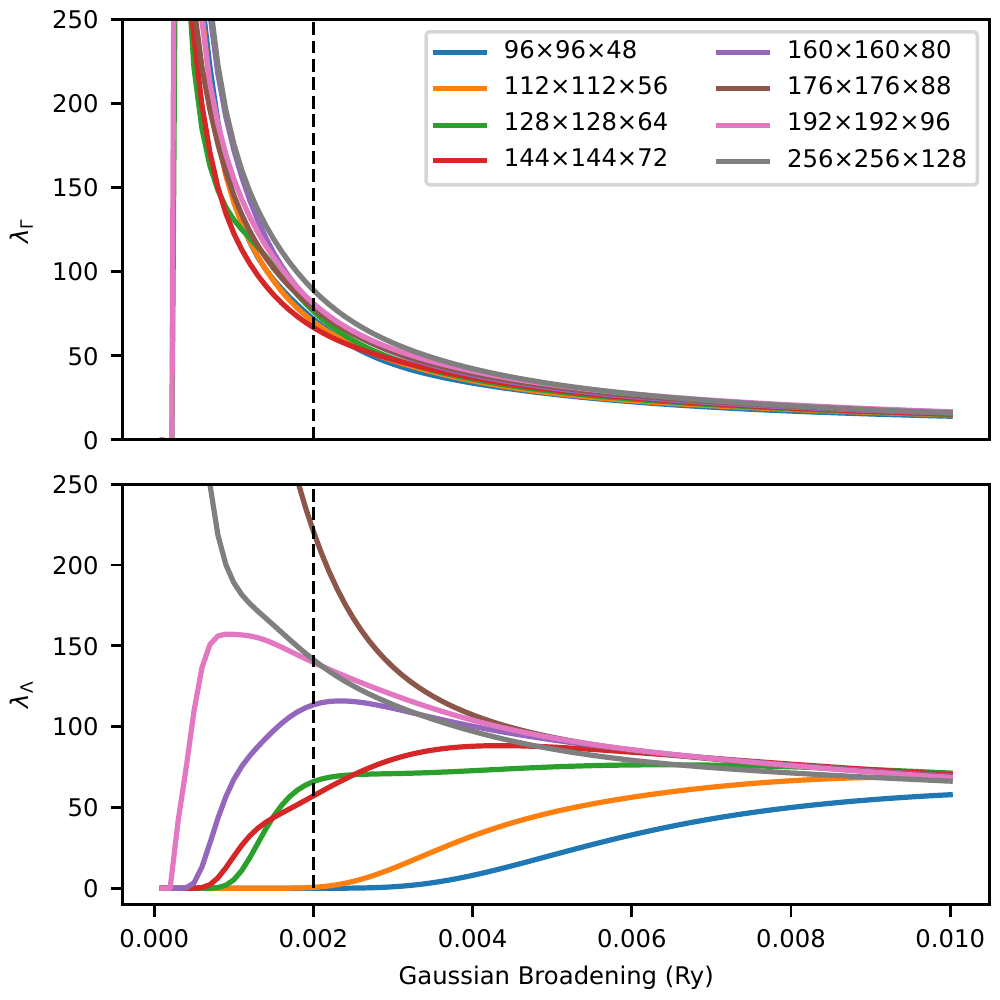}
\caption{Test on the ${\bf k}$-mesh convergence for electron phonon coupling strength at $\Gamma$ and $\Lambda$ point at the pressure of $20$~TPa. It shows that $\Gamma$ point is less sensitive with the ${\bf k}$-mesh. We choose a ${\bf k}$-mesh of 192$\times$192$\times$96 with Gaussian broadening of $0.002$~Ry that guarantee $\lambda$ value converge when comparing with 240$\times$240$\times$120 and 256$\times$256$\times$128.}
\label{fig:kmesh}
\end{figure*}

\newpage

\begin{figure*}[htp]
\centering
\includegraphics[width=0.9\textwidth]{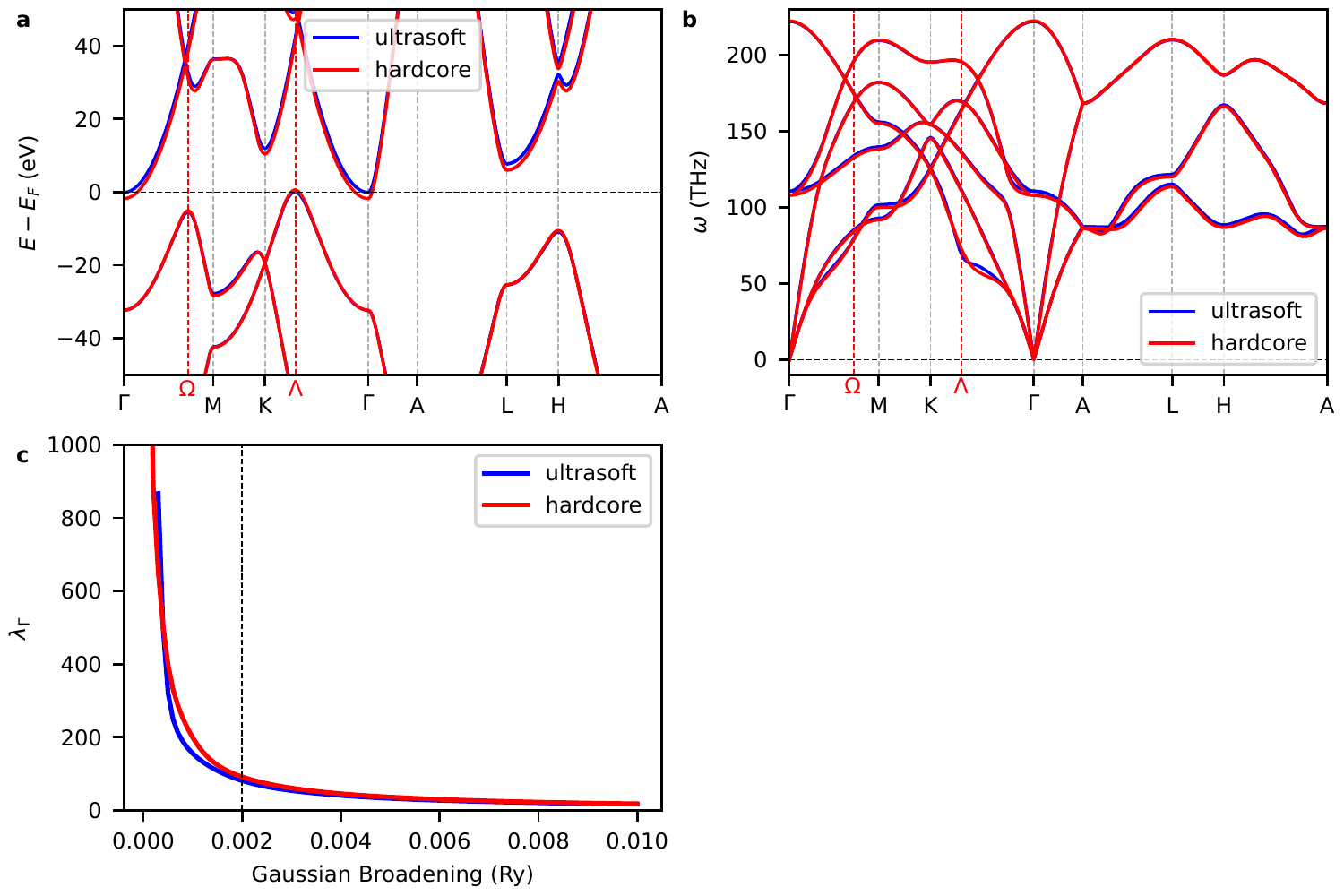}
\caption{Test on the pseudopotentials. We compare ${\bf a}$ bandgap structure, ${\bf b}$ phonon spectrum, ${\bf c}$ electron phonon coupling strength at $\Gamma$ point at $20$~TPa. It is shown that ultrasoft pseudopotentials are reliable for conducting calculations in the TPa regime.}
\label{fig:hardcore}
\end{figure*}

\newpage

\begin{table*}[htp]
\caption{Test on the ${\bf q}$-mesh convergence for $\omega_{log}$ and $T_{c}$ with Coulomb pseudopotential $\mu^{*} = 0.10$ at a pressure of $50$~TPa.}
\label{table:qmesh}
\vspace{0.5cm}
\centering
\begin{tabular}{c c c c}
\hline
\hline
\quad ${\bf q}$-mesh \quad & \quad $\lambda$ \quad & \quad $\omega_{log}$ (K) \quad & \quad $T_c$ (K) \quad \quad  \\
\hline
    6$\times$6$\times$3 & 1.45 & 6300 & 695 \\
    8$\times$8$\times$4 & 1.75 & 5047 & 658 \\
    12$\times$12$\times$6 & 0.33 & 5772 & 7 \\
    14$\times$14$\times$7 & 0.39 & 5441 & 21 \\
    16$\times$16$\times$8 & 0.52 & 5203 & 72 \\
% 18$\times$18$\times$9 & 0.41 & 5237 & 25 \\
% 20$\times$20$\times$10 & 0.34 & 5226 & 7 \\
\hline
\hline
\end{tabular}
\end{table*}

\newpage

\end{document}